

\magnification = 1200
\hsize = 15 truecm
\vsize = 23 truecm
\baselineskip 20 truept
\voffset = -0.5 truecm
\parindent = 1 cm
\overfullrule = 0pt

\def\u{1\kern -0.7em \hbox {1}}
\def\p{\partial}
\def\C{\bf C}
\def\Z{\bf Z}
\def\h{\hbar}

\def\sqr#1#2{{\vcenter{\vbox{\hrule height.#2pt
              \hbox{\vrule width.#2pt height#1pt \kern#1pt
                   \vrule width.#2pt}
            \hrule height.#2pt }}}}
\def\square{\mathchoice\sqr54\sqr54\sqr{2.1}3\sqr{1.5}3}

\def\ms{\mathop {res}\limits_{p=s}}
\def\mi{\mathop {res}\limits_{p=\infty}}
\def\mk{\mathop {res}\limits_{p\in {\rm Ker}\ W'}}

\rightline {May, 1995}

\vskip 1cm

\centerline
{\bf Topological Landau-Ginzburg theory with a rational potential}
\centerline
{\bf and the dispersionless KP hierarchy}

\vskip 1cm

\centerline {\ Shogo Aoyama$^1$ and Yuji Kodama$^2$
\footnote*{Supported in part by NSF grant DMS-9403597}}

\vskip 0.5cm

\centerline {\it $^1$ Department of Physics, Shizuoka University}
\centerline{\it Shizuoka  422,  Japan}
\centerline{\it E-mail: spsaoya@ipcs.shizuoka.ac.jp}

\vskip 0.3cm

\centerline{\it $^2$Department of Mathematics, Ohio State University}
\centerline{\it Columbus, OH \ 43210,  USA}
\centerline{\it E-mail: kodama@math.ohio-state.edu}

\vskip 1cm

\noindent
{\bf Abstract}

Based on the dispersionless KP (dKP) theory,
we give a comprehensive study of the topological Landau-Ginzburg (LG) theory
characterized by a rational potential. Writing the dKP hierarchy in a general
form, we find that the hierarchy naturally includes the dispersionless
(continuous) limit of Toda hierarchy and its generalizations
 having finite number of primaries.  Several flat solutions
 of the topological LG theory are obtained in this
formulation, and are identified with those discussed by
Dubrovin. We explicitly construct gravitational descendants
for all the primary fields. Giving a residue formula for the 3-point
functions of the fields, we show that these 3-point functions satisfy the
topological recursion relation.  The string equation is obtained as the
generalized hodograph solutions of the dKP hierarchy, which show
that all the gravitational effects to the constitutive equations (2-point
functions) can be renormalized into the coupling constants in the
small phase space.

\vfill\eject

\footline={\hss\tenrm\folio\hss}

\noindent
{\bf \S1.~  Introduction }

\vskip 0.3cm

The equivalence of the multi-matrix model to the $2$-d gravity
coupled with comformal matter [KPZ, DK, Da] was shown by solving the
generalized
KdV hierarchy [BKa, DS, GM, BDKS, CGM, Do, BDSS].
 Soon later the $2$-d topological field theory (TFT) coupled to topological
  gravity was constructed, which encodes
 the essential features of these three disciplines, i.e., local coordinate
independence, scaling property and integrability [W].
It was then  shown that the multi-matrix model is equivalent to
 the topological conformal matter system coupled to the topological
 gravity [DW, DVV2].
The TFT coupled to the gravity can be studied in a purely topological
 approach based on the cohomology of the physical observables .
  This cohomological nature  is inherent also
in the minimal conformal matter system and the generalized KdV hierarchy, i.e.
fusion rules and polynomial ring (in the dispersionless limit) respectively.

\medskip

In this regard the Landau-Ginzburg (LG) theory, which had already been useful
in the classification of the conformal field theory [VW], also came close to
the arena of these subjects. Namely, through the correspondence of
the LG potential to the Lax operator of the generalized KdV hierarchy,
it gave a most convenient picture for understanding
the ring structure of the hierarchy (in the dispersionless
limit), i.e. a deformation of the ring by the gravitational couplings.
In fact, genus-zero correlation functions of an arbitrary TFT
 coupled to 2-d topological gravity are determined by
an appropriate LG potential (LG picuture).

\medskip

Among these theories the TFT coupled to the gravity in [W] is most general,
dealing not only with non-conformal matter, but also the geometry
of the matter system.  Of course, the KdV hierarchy can be further
 generalized to the KP hierarchy with a certain reduction, but it is not clear
  how the generalized theory comes across the latter subject.

\medskip

 A coupled system of the  gravity and topological matter fields  $\phi^\alpha$
 (primaries)  is given by the action
 $$
 S = S_0 + \sum_{\alpha \in {\rm primaries} \atop N \ge 0}
          t_{N,\alpha}\int_\Sigma \sigma_N (\phi^\alpha),           \eqno (1.1)
 $$
with $S_0$ the basic action obtained by twisting an ordinary model,
$\sigma_N (\phi^\alpha)$  the $N$th gravitational
desecndant of $\phi^\alpha$, and $t_{N,\alpha}$ the coupling constants.
It is essentially characterized by genus-zero 3-point functions
$<\phi^\alpha\phi^\beta\phi^\gamma>$.  The TFT is then defined such that, when
the descendants couplings are turned off at certain values, the 3-point
functions  $<\phi^\alpha\phi^\beta\phi^\gamma> :=
c^{\alpha\beta\gamma}$ of the theory satisfy a set of equations,
$$
\eqalignno{
\eta^{\alpha\beta} & :=c^{\alpha\beta 0}  =  {\rm constant},
 \quad \quad ({\rm flatness\ of\ metric})   & (1.2a) \cr
 & {} \cr
{\partial \over \partial t_\delta}c^{\alpha\beta\gamma} & =
 {\partial \over \partial t_\alpha}c^{\delta\beta\gamma},
  \quad \quad \quad \quad \quad \quad ({\rm integrability})     & (1.2b)  \cr
  & {}  \cr
\sum_{\lambda, \mu = {\rm primaries}}c^{\alpha\beta\lambda} \eta_{\lambda\mu}
c^{\mu \gamma\delta} & =
\sum_{\lambda, \mu = {\rm primaries}}c^{\alpha\gamma\lambda} \eta_{\lambda\mu}
c^{\mu \beta\delta} \ . \quad  ({\rm associativity})
  & (1.2c)  \cr}
$$
Here $t_\alpha := t_{0,\alpha}$ are the primary couplings in (1.1), and
$\eta_{\lambda\mu}$ is the inverse of $\eta^{\lambda\mu}$.
The set of these equations (1.2) is called the WDVV equation [Du1], and
the solutions of this equation describe the topological limits of the TFT
 coupled to the gravity (the ``flat'' solutions).
In this paper, we call the space of primary couplings $t_{\alpha}$ alone
the ``small'' phase space, and the entire coupling space the ``large'' phase
space.

\medskip

The genus-zero limit of certain TFT's coupled to the gravity is equivalent
to the dispersionless limit of the KP (dKP) hierarchy with a  reduction.
 In [DVV1] the flat solutions to the A-, D-, and E- (ADE-) models of the
 topological minimal models were obtained
by the LG approach.  In particular,
  a structure of the dKP hierarchy was found for the case of the A-model, where
the fusion ring of the primaries is given by a ``polynomial" ring with
one variable $p$, the quasi-momentum, over an ideal given by the LG potential.
The approach based on the dKP hierarchy was further extended in [AK, Kr2, T]
by introducing a ``rational" LG potential, which includes the D-model
and its extention [T].  TFT's with a rational potential
appear in several literatures, such as a classical limit of the multi-matrix
model
[BX], the multi-field
representations of the KP hierarchy [ANPV], and the symmetry constrained KP
hierarchy [OS]. (They are all equivalent, and have the same Lax operators.)

\medskip

A classification of the flat solutions of the WDVV equation was given in [Du2],
 by studying the Frobenius manifold, but not in the dKP approach.
 The group theoretical structure behind the solution was understood
by associating the scale dimensions of the primaries with the degrees of
the Coxeter group. It was then shown that all the classified solutions fall
 into the ADE-series or the relatives by some truncations [Z].
Quite recently the solution of the $CP^1$ model [W, DW] was also found
in the LG approach [EY]. Of course, there are other flat solutions [W, DW]
for which the LG description  is not yet  known.

\medskip

All these flat solutions are obtained as the topological limits of the TFT
coupled to the gravity.
The generalizations of the analysis to the large phase space
  has been carried out in a rather limited  extent, except in [AK, Kr2, LP].
The framework based on the cohomology in [W, DW] provides a perturbative method
to study the TFT in the large phase space.
 But the integrable structure behind such a solution of the theory remains
 obscure. In this regard, the approach based on the dKP hierarchy with
the LG picture prevails that by the cohomological approach.

\medskip

In this paper we give a comprehensive study of the topological LG theory
 having a rational potential based on the dKP hierarchy previously proposed in
[AK].  The paper is organized as follows: In Section 2, we give a mathematical
  background of the dKP theory.  The main purpose in this section is to
  rewrite the dKP hierarchy in a general form, and show that it includes
several fusion rings for a single rational potential.
In particular, the dispersionless
  limit of Toda (dToda) hierarchy [TT1, Ko1, SV] is shown to be naturally
included in this formulation.

\medskip

Section 3 introduces the universal coordinates
for the deformation variables of the rational potential.
Then writing the dKP hierarchy with these coordinates, we show the
integrability of the hierarchy. The generalized Gel'fand-Dikey potentials
in the hierarchy are also given explicitly in a residue formula.
We also find several flat solutions corresponding to the topological limits of
our TFT model.

\medskip

In Section 4, based on the integrability of the hierarchy,
 we define the free energy and the N-point functions of the TFT.
The 3-point function is explicitly expressed by a residue formula.
We also define another type of the free energy and the corresponding N-point
functions, which are the functions of the universal coordinates.  But the
2-point functions of both types are found to be a unique object.
Namely these 2-point functions calculated from the respective free energies
coincide.  Moreover, the 2-point functions are free from the choice of the
primary rings obtained in the general formulation of the dKP hierarchy.
This leads to a symmetry of the WDVV equation discussed in [Du2].
A hamiltonian form for the dKP hierarchy is also found in
 terms of 1-point functions which are also given explicitly.

\medskip

 Section 5 defines the gravitational descendants for all the primary fields.
 In particular, we elaborate the descendants of the primary $\phi^{-1}$
 which is a typical flow in the dToda hierarchy.  The descendant fields of
 $\phi^{-1}$ for the dToda hierarchy (CP$^1$-model) were first found in [EY],
but
 in a rather heuristic way.  Here we give a mathematical justification to
 their result.  Then we show that the
 topological recursion relation for these descendants naturally inheres in
 the framework of the dKP theory.

\medskip

In Section 6, we show that the solution of
the dKP hierarchy is completely determined in the small phase space.
This implies that all the gravitational effects to the constitutive equations
(2-point functions) can be renormalized in the primary couplings.
This is precisely the theorem obtained in [KG], and the string equation
 is then obtained as a consequence of this theorem.
The solution of the string equation is given algebraically in the generalized
hodograph transform, and can be explicitly obtained
as a perturbative solution with small gravitational couplings.
This is the well-known procedure of the renormalization in
the quantum field theory.

\medskip

In Section 7, we discuss the critical phenomena based on our LG theory.
The main objective here is to study the scaling behavior of the solution of the
 dKP hierarchy in the small phase space.  The critical exponents of the free
 energy and the primaries are calculated for the scaling models, which has no
 scaling violation term, such as log-solution.
Among the LG theories having different type
  of rational potentials, we find that there exists an equivalent pair in the
 sense that two theories in this pair give the same scaling behavior
at all the critical points. We also discuss a phase transition corresponding to
 a singularity (shock formation) appearing in the string equation.
The singularity may be regularized by adding a finite genus effect to
the dKP hierarchy  (the Whitham approach) [BKo, Du1, Kr2].

\medskip

In Section 8, we illustrate the results obtained in
this paper by taking several explicit examples including the CP$^1$-model
[DW, EY], and especially we give the corresponding free energies for
the flat solutions.  We also provide a detail analysis of the terms
including log-singularity in Appendix A, and a brief overview of
the dKP theory in Appendix B.

\medskip

In this paper, we restrict ourself to the analysis of the TFT in the
genus-zero limit, which corresponds to a spherical approximation of the
world sheet in the string theory.  Effects of finite genus to the world sheet
may be studied by replacing the quasi-momentum $p$ in the rational potential
by a differential symbol $\p$ (i.e. the multi-matrix models of [BX, Da]),
or by promoting the potential into
a matrix form [KO].   Quantization  of the dKP theory may be also
obtained by the Moyal deformation [S].  We will study these in a future
communication.

\vskip 2cm

\noindent
{\bf \S2.~  Preliminary on the dispersionless KP hierarchy  }

\vskip  0.5cm

In this paper we study a topological Landau-Ginzburg (LG) theory with a
rational potential given by a Laurent polynomial
of $p$ and $(p-s)^{-1}$ [AK, Kr2],
$$
\eqalign{
W & = W(p \ ;v,s)  \cr
& = {1 \over n+1}p^{n+1}  + v_{n-1} p^{n-1} +  \cdots\cdots + v_0   \cr
& \quad \quad \quad   + {v_{-1} \over p-s} + \cdots\cdots + {v_{-(m-1)} \over
(m-1)(p-s)^{m-1}} +    {v_{-m} \over m(p-s)^m} \ .  \cr}  \eqno (2.1)
$$
Here the variable $p$ represents the quasi-momentum of the field, and
 the (complex) coefficients $v_a,\ -m  \le a \le n-1$ and $s$ are
the deformation variables of the potential.
 At the singularities $p = \infty$ and $p = s$ of the W potential we introduce
 the local coordinates $\lambda$ and $\mu$,
$$
\eqalignno{
\lambda &  = p + O\left({1 \over p}\right),
\quad \quad \quad {\rm for \ \ large}\quad p ,  & (2.2a) \cr
 & {} \cr
\mu     & = {\root m \of {v_{-m}} \over p-s} + O(1),
\quad \quad \quad {\rm for \ \ small }\quad  p-s,  & (2.2b) \cr}
$$
which are also globally defined through [AK, T]
 $$
 W =  {\lambda^{n+1} \over n+1} = {\mu^m \over m}.    \eqno (2.3)
$$

\vskip 0.5cm

The main objective in this paper is to study  deformation of
the W potential in the framework of the LG theory
for TFT, where the deformation is induced by  coupling constants.
Here we introduce an infinite number of coupling  constants
 $t_i$ for $\ i \in {\rm \Z} \ $, and consider the deformation variables
 $\{v_a\}$ and $s$ to be functions of $t_i$'s. Then we assume the W potential
 to satisfy the following flow equations of the deformation,
$$
{\partial W \over \p t_i} = \{Q^i, W \}
                       := {\p Q^i \over \p p}{\p W \over \p t_0}
                        - {\p Q^i \over \p t_0}{\p W \over \p p}, \quad
  \quad  \quad {\rm for} \quad i \in { \ \rm \Z} \eqno (2.4)
$$
with the genrators $Q^i$  defined by
$$
Q^i  := \left\{
 {\matrix{ \left[ {{1 \over i+1}\lambda^{i+1}} \right]_+,
 \quad \quad \quad \quad \quad {\rm for}\quad 0 \le i < \infty,     \cr
                {} \cr
[\ \log \lambda \ ]_+ -[\ \log \mu \ ]_-,    \quad \quad \quad {\rm for}
\quad i \ = -1 \ ,   \cr
{} \cr
 \ \  -\left[ {{1 \over |i|-1}\mu^{|i|-1}} \right]_- ,
    \quad \quad \quad   {\rm for}\quad - \infty < i \le - 2 .    \cr
            }} \right.     \eqno (2.5)
$$
The symbols
$[\cdot]_+$ and $[\cdot]_{-}$ indicate the parts of non-negative powers in $p$
  and negative powers in  $p-s$, respectively.  In Appendix A, we give
  a precise definition of
  these $\pm$ projections, and also provide explicit caluculations
  of the terms including the $\log$ - terms, $\log \lambda$ and $\log \mu$.
  In particular, one can show that $Q^{-1} = \log (p-s)$, which was previously
  used in [AK, Kr2].
  It should be also noted that the definition of $Q^{-1}$ in (2.5) is
natural, even
  though it looks complicated.  Indeed, this definition clearly shows that
  $Q^{-1}$ is a generator of degree zero in $p$ having contributions from
  both singularities $p=\infty$ and $p=s$ in the rational potential $W$.

\medskip

Eq. (2.4) is nothing but the dispersionless KP (dKP) hierarchy
[KG, TT2, Kr1, Du1] with the reduction given by the W potential (2.1). We
refer the set of eqs.(2.4) the dKP hierarchy in this paper, even though we
mainly concern
with the reduced one. In Appendix B, we briefly summarize the theory of the dKP
 hierarchy.  Note here that each deformation in (2.4) should be
 independent from the others.
Namely we have to have the compatibility conditions
 among the flows in (2.4).  In order to confirm the conditions, we fisrt note:
\proclaim Lemma 2.1.
The generators $Q^i$ satisfy the zero curvature condition,
$$
{\p Q^i \over \p t_j} - {\p Q^j \over \p t_i} + \{Q^i, Q^j\} = 0, \quad
\quad {\rm for} \quad   i, j \in { \ \rm \Z} \ . \eqno (2.6)
$$

\noindent
{\it Proof.}~~~
Here we give a proof in the general form:  Let $F$ and $G$ be the functions of
either $\lambda$ or $\mu$, and define $F_{\pm}:=[F]_{\pm}$
and $G_{\pm}:=[G]_{\pm}$.
The flow parameters corresponding to
$F_{\pm}$ and $G_{\pm}$ are denoted by $t_{\pm}$
and $s_{\pm}$, respectively.  Then, for the case with $F_+$ and $G_-$, we have
$$
\eqalign{
{\p F_+ \over \p s_- }
& = \left[ {{\p F \over \p s_-}} \right]_+
 = \{G_- \ ,F\}_+  \cr
& = \{G_- \ ,F_+\} - \{G_-\ , F_+\}_- \cr
 & =
\{G_-\ ,F_+\} + {\p G_- \over \p t_+}.    \cr}
$$
For the case with $F_+$ and $G_+$, that is, both $F$ and $G$ are the
functions of $\lambda$, we have
$$
\eqalign{
{\p F_+ \over \p s_+ }
& = \left[ {{\p F \over \p s_+}} \right]_+
 = \{G_+\ , F \}_+ \cr
& = \{G_+\ ,F_+ \} + \{G_+\ , F_- \}_+  \cr
 & =  \{G_+\ , F_+ \} + \{G\ , F_- \}_+  \cr
& =
\{G_+\ ,F_+ \} - \{G\ , F_+ \}_+  \cr
& =
\{G_+\ ,F_+\} + {\p G_+ \over \p t_+}.    \cr }
$$
Using these results, the case including $Q^{-1}$ can be also shown by a
similar but careful computation (see Appendix A for a detail).$\quad \square$

\vskip 0.3cm

\noindent
We then  obtain:
\proclaim  Proposition 2.2. The flows in eq. (2.4) are compatible
(or commuting) for all $t_i$'s,
$$
{\p \over \p t_i}{\p \over \p t_j} W = {\p \over \p t_j}{\p \over \p t_i} W .
\eqno (2.7) $$

\noindent
{\it Proof.}~~~~
{}From eq. (2.4) the compatibility condition can be written in the form
$$
\eqalign{
{\p \over \p t_i}{\p \over \p t_j}W - {\p \over \p t_j}{\p \over \p t_i} W
& = \left\{ { {\p Q^j \over \p t_i} - {\p Q^i \over \p t_j} + \{Q^j, Q^i\} \ ,
 W  }  \right\} \ .
 \cr}
$$
Use of Lemma 2.1 completes the proof. $\quad \square$

\vskip 0.5cm

\noindent
We thus see that the flows commute if those generators are the functions of
$\lambda$ or/and $\mu$ only.
Proposition 2.2 implies that the dKP hierarchy (2.4) possesses
an infinite number of symmetries inducing the conservation laws
(see Theorem.3.4),
and leading to its integrability.  This gives the main reason
why we use the dKP hierarchy to express the deformation with an infinite number
of coupling constants.

\medskip

Let us now consider a ring of Laurent polynomials  over $\C$,
 denoted by  ${\C} [p, (p-s)^{-1}]$. A basis for this ring may be given by
$$
\phi^i := {d Q^i \over dp}, \quad \quad \quad  i \in \Z \ . \eqno (2.8)
$$
Namely $\phi^i$ is a Laurent polynomial of degree $i \in \Z \ $.
Introducing an ideal given by
$$
{d W \over d p} := W' = \phi^n - \phi^{-(m+1)} = 0,  \eqno (2.9)
$$
we have a (commutative and associative) finite dimensional rational
ring of dimension $n+m+1$,
$$
{\cal R} := {{\rm \C}[p, (p-s)^{-1}] \over W'(p) } .   \eqno (2.10)
$$
A basis of this ring ${\cal R}$ can be taken to be a set of
$n+m+1$ consecutive
elements in eq. (2.8), including $\phi^0$ as an identity element of the ring.
We thus consider here a topological field
theory with $n+m+1$ primary fields.
Since $\phi^n$ and $\phi^{-(m+1)}$ are identified by eq. (2.9),
 a natural basis may be chosen as
$$
\{ \phi^\alpha : \alpha \in \Delta_{n,m} \},    \eqno (2.11)
$$
where the set of indicies $\Delta_{n,m} \subset \Z$ is given by
$$
\Delta _{n,m}:=\{ i \in {\Z}  : -m\le i\le n\}.  \eqno (2.12)
$$
The fields $\phi^\alpha$ in (2.11) are called the ``primaries", which
describe the matter fields of our TFT,
while the other $\phi$'s  the ``gravitational descendants".
(Throughout this paper,
we use the Greek letters for the primary indices, and the Roman letters for
all indices including both primary and descendants indices.)
The TFT is then described by an action,
$$\eqalign{
S &= S_0 \ + \ \ \sum_{i \in \Z} t_i \phi^i  \cr
&= S_0 \ + \sum_{\alpha \in \Delta_{n,m}}
t_{\alpha} \phi^{\alpha} \ + \sum_{i \in {\Z} \backslash \Delta_{n,m}}
t_i\phi^i , \cr} \eqno (2.13)
$$
where $S_0$ is the matter sector action.  In this formula, the
coupling constants $t_\alpha$ with $\alpha \in \Delta_{n,m}$ are
called the primary
couplings describing the deformation of the matter sector of the TFT, while
all the others the gravitational couplings describing the gravitational
deformations.

\medskip

 In the view point of the dKP hierarchy (2.4), the fields $\phi^i$ and
the coupling constants $t_i$ are related by
$${{\partial W(p_*)} \over {\partial t_i}}
=\phi ^i(p_*){{\partial W(p_*)} \over {\partial t_0}},  \eqno (2.14)
$$
where $p_*$ is a root of the ideal $W'(p)=0$, denoted by
$p_* \in $ Ker$W'$. This is the Riemann invariant form of the flow equation
(2.4) which is a quasi-linear system
of first order equations for the deformation variables $v_a$ and $s$ in $W$.
In this form of the dKP hierarchy, we note that the primary
coupling $t_0$ plays a particular role, called the ``cosmological constant"
in the string theory, and the corresponding field is the identity $\phi^0$.
These are customarily denoted as $t_0=t_{\cal P}$ and $\phi^0= {\cal P}$,
where ${\cal P}$ is called the ``puncture" operator.  However,
writing the dKP hierarchy (2.4) in a more general form, one can define a
different set of primary fields, and associate $t_{\cal P}$ with an other
primary coupling  $t_{\alpha_0}$ for some $\alpha_0 \ne 0$, whose field is
of course the identity of this set.
This general form of the dKP hierarchy may be given by
$$
\{ Q_j \ , W \}_i = \{ Q_i \ , W \}_j ,  \eqno (2.15)
$$
where $\{A \ , B\}_i$ represents the Poisson bracket with $(p, t_i)$ pair, i.e.
$$
\{A,B\}_i:={{\partial A} \over {\partial p}}{{\partial B} \over {\partial t_i}}
-{{\partial A} \over {\partial t_i}}{{\partial B} \over {\partial p}}.
\eqno (2.16)
$$
This is derived from (2.4) and (2.6), and the dKP hierarchy
in (2.4) corresponds
to the case with $j=0$.  From the form (2.15) with fixed $j=\alpha_0
 \in \Delta_{n,m}$, the flow equation for the W potential similar to (2.14)
takes the form,
$$
{{\partial W(p_*)} \over {\partial t_i}}=
{ {\tilde \phi }^{i}(p_*)}{{\partial W(p_*)} \over {\partial t_
{\alpha_0} }}
:={{\phi ^i(p_*)} \over {\phi ^{\alpha_0} (p_*)}}
{{\partial W(p_*)} \over {\partial t_{\alpha_0}}}.  \eqno (2.17)
$$
This defines a set of new primary fields with a fixed
$\alpha_0 \in \Delta_{n,m}$,
$$
\{ \ {\tilde \phi}^{\alpha} :=
{ \phi^{\alpha} / \phi^{\alpha_0}} \ \ ({\rm mod} \ W') \ : \
\alpha \in \Delta_{n,m} \ \},  \eqno (2.18)
$$
whose identity element is given by  $\tilde \phi^{\alpha_0}$. This implies that
we have $n+m+1$ different choices of the puncture operator, ${\cal P} =
\tilde \phi^{\alpha_0}$, and the cosmological constant $t_{\cal P}
= t_{\alpha_0}$. Note that each new field ${\tilde \phi}^{\alpha}$ is also
defined as an element of the ring ${\cal R}$ in (2.10).  One should  also note
that the hierarchy (2.15) is mathematically equivalent to (2.4),
and the solutions of
this hierarchy are of course the same as those of (2.4).  However, as
we explained above, the physical significance of the flow parameters
$t_\alpha$ is different, and this observation will be useful to construct
various solutions relevant to our TFT.
In particular, the dKP hierarchy in (2.15) with $j=\alpha_0=-1$ turns out to be
the dispersionless (or continuous) limit of the generalized Toda
hierarchy [Ko1], and introducing $P:=p-s$, (2.15) becomes
$$
{\p W \over \p t_i} = P \left( {{\p Q_i \over \p P}{\p W \over \p t_{-1}}
-{\p Q_i \over \p t_{-1}}{\p W \over \p P} } \right).  \eqno (2.19)
$$
The right hand side gives the Poisson bracket for the dispersionless
Toda hierarchy [TT1].
The basis of this ring is then given by
$$
\{{\tilde \phi}^{\alpha}=P \phi ^{\alpha} \
({\rm mod} \ W') \ : \alpha \in \Delta_{n,m}\}\
\subset {\C}[P,P^{-1}],
$$
and the W potential is
$$
W={1 \over {n+1}}P^{n+1} + w_nP^n + \cdots + w_0 + {w_{-1} \over P} +
{w_{-2} \over 2P^2} + \cdots + {w_{-m} \over mP^m}.    \eqno (2.20)
$$
\vskip 0.3cm

\medskip

 With the ideal (2.9),
the fusion algebra on the ring ${\cal R}$ is defined  by the product rule,
$$
\phi^{\alpha} \phi^{\beta} = \sum_{\gamma \in \Delta_{n,m} \ }
c^{\alpha\beta}_{\;\gamma}\phi^\gamma
\quad ({\rm mod}\ \ W'), \quad {\rm for } \quad
\alpha, \beta \in \Delta_{n,m} \ , \eqno (2.21)
$$
with the structure constants $c^{\alpha\beta}_{\;\gamma}$.
Associativity of the fusion algebra plays a fundamental role for the
TFT described by the W potential. It is then an important subject  to study
the structure constants as functions of  the coupling constants $t_i$ for
$ i \in \Z\ $.
 We will study the fusion algebra (2.21) in terms of the 3-point functions
 in Section 4.

\medskip

As a final remark of this section, we note that our choice of the rational
ring on ${\C}[ p, (p-s)^{-1}]$
can be naturally extended to a more general case with multi-poles at
$p=s_i, i=1, \cdots ,l$, proposed in [ANPV, BX, Kr2].  In fact, because of
asymmetric form of $p$ and $(p-s)^{-1}$, it is immediate to see that
a ring in this general case is defined on the Laurent polynomials
in ${\C}[p, (p-s_1)^{-1}, \cdots , (p-s_l)^{-1}]$.

\vskip 2cm

\noindent
{\bf \S3.~ The dKP hierarchy in the universal coordinates}

\vskip 0.5cm

The dKP hierarchy (2.4) defines the  flows of the variables
$v_a, \  -m \le a \le n-1$ and $s$. In this section we reformulate
the dKP hierarchy (2.4) in terms of new variables, that is, a reparametrization
of the deformation variables.  These new variables which we refer as
the ``universal coordinates" are introduced as follows: First  we invert
 eqs. (2.2a) and (2.2b) in terms of $p$ respectively
(see also Appendix B for a motivation of this procedure),
$$
\eqalignno{
p & = \lambda - {u^0 \over \lambda} -
 \cdots\cdots\cdots -  {u^i \over \lambda^{i+1}} - \cdots,
 \quad \quad {\rm for\ large}\ \lambda,   & (3.1a) \cr
p & = u^{-1} + {u^{-2} \over \mu} +
 \cdots\cdots\cdots +  {u^{-i} \over \mu^{i-1}} + \cdots,
 \quad {\rm for\ large}\ \mu.  & (3.1b) \cr}
$$
 Then we  have:
\proclaim Lemma 3.1. The coefficients  $ u^i$ for $  i \in {\Z}$
  can be expressed as the residue formulae,
$$
u^i  = \left\{ {\matrix{{1 \over i+1}\mi [\lambda^{i+1}],\quad \quad \quad
  {\rm for} \quad  0 \le i < \infty,   \cr
  {}  \cr
 \mi [\ \log \lambda \ ] + \ms [ \ \log  \ \mu \ ],  \quad
 {\rm for} \quad i \ =\ -1, \cr
  {}  \cr
 \ \ \ {1 \over |i|-1}\ms [\mu^{|i|-1}],
 \quad \quad \quad {\rm for} \quad - \infty < i \le -2.  \cr} } \right.
\eqno (3.2)
$$
(The residue formula is defined in the usual way as (A.3) in Appendix A.)

\noindent
{\it Proof.} These formulae except $i=-1$ can be shown by replacing
the differential $dp$ in the residue integral with that of
the local coordinate, i.e.
$ dp = (d p / {d \lambda}) d \lambda$ or $  dp = (d p / {d \mu}) d \mu.$
As explained in Appendix A, the $u^{-1}$ is evaluated  as
$$\eqalign{
& u^{-1} = \mi [ \ \log \lambda \ ] + \ms [ \ \log \mu \ ]
= \mi [ \ \log \ p \ ] - \ms [ \ \log \ (p-s) \ ]  \cr
& {} \cr
& = {1 \over {2\pi i}} \left[\oint_{C_\infty} \log p \ dp -
 \oint_{\tilde C_s} \log (p-s)  \ dp \right]
  = {1 \over {2\pi i}}\oint_{C_\infty} \log \left({p \over p-s }
 \right) \ dp
= s, \cr }
$$
where the contour $C_\infty$ is taken arround $p=\infty$, and
$\tilde C_s$ to surround a branch cut between $p=s$ and $p=\infty$, in both
the directions of counter-clockwise. $ \quad \square$

\medskip

Note in (3.2) that the coefficients $u^{\alpha}$ for $\alpha \in \Delta_{n,m}$
are  determined from the deformation variables $v_a$ and s in eq. (2.1),
while the others are polynomials of
these  $u^\alpha$ (see Proposition 3.3 for their explicit forms).
These variables $u^\alpha$ play an important role throughout this paper, and
we call them the universal coordinates of the deformation.
One of the main purpose
of this paper is to construct them as functions of the coupling constants
$t^i$ by solving the dKP hierarchy (2.4).

\medskip

We note  that the universal coordinates $u^\alpha$ are related to the
primaries $\phi^\alpha$ through the W potential:
\proclaim Proposition 3.2.
For each primary index $\alpha \in \Delta_{n,m}$, we have
$$
{\p W \over \p u^\alpha} = \phi_{\alpha}
:= \sum_{\beta \in \Delta_{n,m}} \eta_{\alpha\beta}\phi^\beta,
\quad \quad ({\rm mod} \ W') \eqno (3.3)
$$
where a metric $\eta_{\alpha\beta}$ is defined by
$$
 \eta_{\alpha\beta}
= \left\{ {\matrix{
\delta_{\alpha+\beta,n-1}, \quad {\rm for}\quad
-1\le \alpha, \beta \le n,  \cr
 {}  \cr
\quad \delta_{\alpha + \beta , -m-2}, \quad  {\rm for} \quad
-m \le \alpha,\beta \le -2,    \cr
{}   \cr
0,\quad \quad \quad  {\rm  otherwise.} \quad \quad \quad \quad  \cr}} \right.
     \eqno (3.4)
$$
(Throughout this paper, we use $\eta_{\alpha\beta}=\eta^{\alpha\beta}$ for
lowering and raising the primary indices.)
\vskip 0.5cm

\noindent
{\it Proof.}~~~ We differentiate  $u^\alpha$ in (3.2) with respect to
$u^\beta$ to find  for $0 \le \alpha,\beta \le n$,
$$
\delta_{\alpha,\beta} = \mi \left[{\lambda^\alpha{\p \lambda \over \p u^\beta}
} \right] = \mi \left[{\lambda^{\alpha-n}{\p W \over \p u^\beta}} \right].
$$
On the other hand the definition of $\phi^\beta$ in (2.8) leads to
$$
\mi[\lambda^{\alpha-n}\phi^\beta] =
\mathop {res}\limits_{\lambda = \infty}
   [\lambda^{\alpha-n}\lambda^\beta ] =
  \delta_{\alpha+\beta,n-1}  \eqno (3.5)
$$
for $0 \le \beta \le n$.  By inspecting (3.1) we note for $0 \le \beta \le n$
$$
{\p W \over \p u^\beta} \in {\rm \C} [p].
$$
Then the uniqueness of  polynomials which are orthogonal to
$\lambda^{\alpha-n}$ gives
 eq. (3.3) for $ 0 \le \alpha \le n$. A similar calculation leads to the cases
 for $-m \le \alpha < 0$.$ \quad \square$

\vskip 0.5cm

\noindent
The formula (3.3) is a Legendre transform between
$$
\{u^\alpha : \alpha \in \Delta_{n,m} \} \quad {\rm  and}\quad
\{\phi^\alpha : \alpha \in \Delta_{n,m} \}
$$
with the generator $W$, i.e. $dW=\sum_{\alpha \in \Delta_{n,m}}
 \phi^{\alpha} du_{\alpha}$, and gives an inversion formula for a
reconstruction of the W potential  from the universal coordinates
$u^\alpha$. In fact we have:
\proclaim Proposition 3.3.  The variables $v_a,\  -m \le a \le n-1$
 and $s$ can be found as the functions of the universal coordinates
 $u_\alpha := \sum_{\beta \in \Delta_{n,m}} \eta_{\alpha\beta} u^\beta$,
$$
\eqalignno{
v_{-a} & = \sum_{\alpha_1 + \cdots + \alpha_a
= (a-1)m + a \atop \alpha_1,\cdots,\alpha_a > 0}
u_{-\alpha_1}u_{-\alpha_2}\cdots\cdots u_{-\alpha_a},   \cr
 & \quad \quad \quad \quad \quad \quad \quad \quad \quad\quad \quad \quad
   \quad \quad \quad \quad \quad \quad
  \quad  {\rm for}\ \  1 \le a \le m, & (3.6a) \cr
 \ \ \cr
v_a & = u_a + \sum_{b=2}^{n-a-1}
{(\alpha+\beta-1)! \over {ab!}}
  \sum_{ \alpha_1+\cdots +\alpha_b  = (b-1)n+a+b-1 \atop  n-1 \ge
      \alpha_1,\cdots,\alpha_a \ge 0}
 u_{\alpha_1}\cdots\cdots u_{\alpha_b}, \cr
 \ \ \ \cr
 & \quad \quad \quad \quad \quad \quad \quad \quad \quad\quad \quad \quad
   \quad \quad \quad \quad \quad \quad
    {\rm for} \ \ 0 \le a \le n-1, & (3.6b) \cr
    \ \ \cr
s & =  u_n.  & (3.6c) \cr}
$$

\noindent
{\it Proof.}~~~  Through the differentiation of the W potential by
the universal coordinates
 we obtain for $a, \alpha \ge 0$
$$
 \eqalign{
{\p v_{-a} \over \p u_{-\alpha}} & =  \mi [\phi^{-\alpha} (p-s)^{a-1}] , \cr
{\p v_{a} \over \p u_{\alpha}} & =  \mi [\phi^{\alpha} p^{-(a+1)}] , \cr
  {\p v_{-a} \over \p u_{\alpha}} & = {\p v_{a} \over \p u_{-\alpha}} = 0 .
\cr }
 \eqno (3.7)
$$
Further differentiation gives the following recursion relations
$$
\eqalign{
{\p \over \p u_{-\alpha}}{\p \over \p u_{-\beta}}v_{-a}  & =  (a-1)
 {\p \over \p u_{-(\alpha + \beta -m-1)} }v_{-a+1},\quad \quad
(\alpha,\beta \not= 1) ,
  \cr
{\p \over \p u_{\alpha}}{\p \over \p u_{\beta}}v_{a}  & =  (a+1)
 {\p \over \p u_{\alpha + \beta -n } }v_{a+1}.  \cr}   \eqno (3.8)
$$
{}From (3.7) we have
$$
v_{-1}  = u_{-1},  \quad \quad \quad
v_{n-1}   = u_{n-1}, \quad \quad \quad  v_{n-2} = u_{n-2},   \eqno (3.9a)
$$
and
$$
{\p v_{-a} \over \p u_{-1}} = \delta_{a,1}, \quad \quad \quad
{\p v_{a} \over \p u_{n}} = 0,\quad \quad \quad {\rm for}\ -m \le a \le n-1.
 \eqno (3.9b)
$$
Then the formulae (3.6) can be obtained by solving  (3.8) recursively with
initial conditions (3.9a).  $\quad \square$

\vskip 0.5cm

The most important aspect of the coefficients  $u^i$ in (3.1) is that
 they give the conserved densities of the dKP hierarchy. Namely we have:

\proclaim Theorem 3.4. There exist  functions $G^{ij}=G^{ij}(u)$ such that
 $$
 { \p u^i \over \p t_j } = { \p \over \p t_0 } G^{ij}, \quad \quad
  i,j \in \Z. \eqno (3.10)
 $$

\noindent
{\it Proof.}~~~
This can be proved by a general formulation of the dKP hierarchy, which
does not depend on the form of $W$ (see Appendix B).
Here we give a proof by a direct calculation using the
explicit formula of $u^i$ given by (3.2): For $i \ge 0 $ and any $j \in \Z \
$ the quantity $G^{ij}$
can be obtained as
$$
\eqalign{
{\p u^i \over \p t_j } & = \mi {\left[ {\lambda^i{\p \lambda \over \p t_j}}
\right] }
  = \mi [\lambda^i\{Q^j, \lambda\}]  \cr
  & {} \cr
& = \mi \left[ {
{1 \over i+1}\{Q^j, \lambda^{i+1}\}} \right]
 = {\p \over \p t_0} \mi {\left[ {{1 \over i+1}\lambda^{i+1} \phi^j} \right] }.
 \cr }
$$
For other cases, similar calculations leads to the following explicit
formulae for $G^{ij}$:
$$
\eqalignno{
G^{ij} & = {1 \over i+1}\mi [\lambda^{i+1} \phi^j]
= {1 \over j+1}\mi[\lambda^{j+1} \phi^i], &  \cr
& \quad\quad\quad\quad\quad\quad\quad\quad\quad\quad\quad\quad\quad\quad\quad
    0 \le i,j < \infty,   &   (3.11a)  \cr
   &  &  \cr
G^{-i-j} & = \left\{  {\matrix{ {
{1\over i-1} \ms [\mu^{i-1}\phi^{-j}],  \quad
2\le i < \infty, \quad 1  \le j < \infty, } \cr
{} \cr
{{1\over j-1} \ms [\mu^{j-1}\phi^{-i}],  \quad
1\le i < \infty, \quad 2  \le j < \infty,}    \cr}}  \right. & (3.11b) \cr
&  & \cr
& &  \cr
G^{i-j}  & = G^{-ji} = \left\{  {\matrix{ {
{1\over i+1} \mi [\lambda^{i+1}\phi^{-j}],  \quad
0\le i < \infty, \quad 1  \le j < \infty, } \cr
{} \cr
{{1\over j-1} \ms [\mu^{j-1}\phi^i],  \quad
0\le i < \infty, \quad 2  \le j < \infty,}    \cr}}  \right. & (3.11c) \cr
&  & \cr
& &  \cr
G^{-1-1} & = \mi [ \ \log \lambda \ \phi^{-1}] + \ms [ \ \log \mu \ \phi^{-1}]
= \log u_{-m} .    &  (3.11d)  \cr}
$$
These can be more easily obtained by using the formula (4.6)
in Proposition 4.3.
$\square$

\vskip 0.4cm

\noindent
Note from (3.11) that $G^{ij}$ are symmetric in the indices,
$G^{ij} = G^{ji}$,
thereby
$$
{\p u^i \over \p t_j} = {\p u^j \over \p t_i }
  \eqno (3.12)
$$
We also note from (3.2) that
$$
u^i=G^{i0}.  \eqno (3.13)
$$
The densities $G^{ij}$ are referred as the generalized Gel'fand-Dikey (GD)
 potential.  As we will show in the next section,
these expressions of the generalized GD potentials will lead to the
definitions of the N-point functions and the free energy of our TFT.

\medskip

{}From Theorem 3.4, we also have:
\proclaim Corollary 3.5.  The generalized GD potential $G^{ij}$ satisfy

$$
{\p  \over \p t_k} G^{ij} = {\p  \over \p t_i} G^{jk} , \quad \quad {\rm for}
\quad  i, j, k \in {\Z}.  \eqno (3.14)
$$

\noindent
{\it Proof.}~~~  Taking the derivative of (3.10) with respect to $t_k$, and
using the commutativity of the flows in $t_j$ and $t_k$, we obtain
$$
{\p  \over \p t_0}{\p  \over \p t_k} G^{ij} = {\p  \over \p t_0}
{\p  \over \p t_j} G^{ik}. $$
This leads to (3.14), except an integration constant which may be
taken to be zero.$\quad \square$

\vskip 0.3cm

\noindent
The Corollary implies that the $G^{ij}$ can be further integrated by both
$t_i$ and $t_j$.  This fact will be important in the next section where we
define N-point functions from $G^{ij}$ ( Proposition 4.2).

\medskip

The formula (3.14) gives a general form of the dKP hierarchy, and is
in fact equivalent to the general form of the hierarchy (2.15). For each $k =
\alpha_0 \in \Delta_{n,m}$ in (3.14), one can then define a new set
of hierarchy,
$${\p  {\tilde u}^i \over \p t_j} ={\p  \over \p t_{\alpha_0}} G^{ij},
\quad \quad i, j \in {\Z},  \eqno (3.15)
$$
where the new variable ${\tilde u}^i$ is a generalization of $u^i$ in (3.13),
and is defined as
$$ {\tilde u}^i = G^{i\alpha_0}.  \eqno (3.16) $$
Then the Riemann invariant form of the quasi-linear system (3.15) is given by
(2.17).  As we will show in Section 4, the primary fields
${\tilde \phi}^{\alpha}$ in (2.18) are related with the generalized universal
 coordinates
  ${\tilde u}^{\alpha}$ by the same Legendre transform as in (3.3), i.e.
$$
{\p W \over \p {\tilde u}^\alpha} = {\tilde \phi}_{\alpha}
:= \sum_{\beta \in \Delta_{n,m}} \eta_{\alpha\beta}{\tilde \phi}^\beta
\quad \quad ({\rm mod} \ W'). \eqno (3.17)
$$

\medskip

In (3.15) the dKP hierarchy is defined
over the entire phase space of the coupling constants.  However, if we
restrict the hierarchy only on the small phase space, we obtain:

\proclaim Proposition 3.6.  The dKP hierarchy (3.15) with $i,j \in \Delta_{n,m}
$  possesses the following solutions for each $\alpha_0 \in \Delta_{n,m}$,
$$
{\tilde u}^{\alpha} = \sum_{\beta \in \Delta_{n,m}} \eta^{\alpha\beta}
t_\beta,   \quad \quad \quad  \alpha \in \Delta_{n,m}.  \eqno (3.18)
$$
As a special case with ${\alpha_0} =0$, we have $u_\alpha = t_\alpha$.

\noindent
{\it Proof.}~~~ Note from (3.17) and (3.18) that we have
$
 \p W / \p t_{\alpha_0} =  \p W / \p {{\tilde u}_{\alpha_0}}
= {\tilde \phi}^{\alpha_0} = 1 \ ({\rm mod} \ W').
$
Then calculating (3.15) with (3.18) and finding the same equation as (3.5) in
 Proposition 3.2 verifies the assertion. $\quad \square$

\vskip 0.3cm

\noindent
  With the solution
(3.18), the primary coupling $t_{\alpha_0}$ then gives the deformation
parameter (cosmological constant) corresponding to the puncture operator
$\tilde \phi^{\alpha_0} = {\cal P}$ in the new set of primaries
$\{ \tilde \phi^{\alpha} \}$  of (2.18), as we have described in Section 2.

\medskip

It is obvious but important to note that the solution (3.18) do not satisfy
the dKP hierarchy for
the gravitational couplings $t_i$. For each $\alpha_0$ the solution of (3.18)
 gives a ``flat" solution of the WDVV equation (1.2), and
  describes the matter sector of TFT at the topological point [DW] (see
  Section 7).  Proposition 3.6 then implies that the dKP hierarchy
  with a given W potential admits
the same number of flat solutions as of the primaries, that is, $n+m+1$ for
W in (2.1). In order to construct  solutions with the gravitational couplings,
  one needs to solve the dKP hierarchy in the entire phase space.
It is however   surprising that these solutions can be obtained by solving
the dKP hierarchy  only in the small phase space. This has been shown
in [KG], and may be considered as the renormalizability of the universal
coordinates.  This will be further   discussed  in Section 6.

\vskip 2cm

\noindent
{\bf \S4.~ N-point functions}

\vskip 0.5cm

In this section we give a realization of our TFT by constructing explicit
formulae of the N-point functions in the framework of the dKP hierarchy.
Let us begin with:
\proclaim    Definition 4.1.
A complex function $<\phi^{i_1}\cdots\phi^{i_N}>$
of $t = (t_i : i \in \Z )$ is a N-point function of the fields
$\{\phi^i\}$, if there exists a function $F = F(t)$
such that
$$
<\phi^{i_1}\cdots\cdots\phi^{i_N}>(t) = {\p \over \p t_{i_1}}\cdots \cdots
{\p \over \p t_{i_N}} F(t).  \eqno (4.1)
$$
Here the function $F(t)$ is called the free energy of a TFT.

\medskip

\noindent
{}From Theorem 3.4 and Corollary 3.5 it is immediate that:
\proclaim Proposition 4.2. There exists a function $F = F(t)$ such that
 the generalized GD potential $G^{ij}$ in (3.11) is expressed by
$$
G^{ij} = {\p \over \p t_i}{\p \over \p t_j} F.  \eqno (4.2)
$$

\medskip

\noindent
It is then natural to identify  $G^{ij}$ to be
the 2-point function generated by the free energy $F$,
$$
<\phi^i\phi^j> = G^{ij}(u).    \eqno (4.3)
$$
This gives a constitutive equation of the 2-point function as
a function of the universal coordinates [DW].  Also with (4.3), the variables
${\tilde u}^i$ in (3.16) with a fixed $\alpha_0 \in \Delta_{n,m}$
 are expressed by
$$
{\tilde u}^i=<\phi^i \phi^{\alpha_0}>,  \quad \quad \quad i \in {\Z}.
 \eqno (4.4)
$$

Now we define the N-point function of our TFT  by differentiating (4.3)
with respect to the flow parameters,
$$
<\phi^i\phi^j\phi^{k_1}\cdots\cdots\phi^{k_{N-2}}> = {\p \over \p t_{k_1}}
\cdots\cdots{\p \over \p t_{k_{N-2}}}<\phi^i\phi^j> \eqno (4.5)
$$
In particular one can prove:
\proclaim Proposition 4.3. The 3-point function defined by (4.5)
can be represented by the residue formula
$$
<\phi^i\phi^j\phi^k> = \mk \left[{{\phi^i\phi^j\phi^k {\p W\over \p t_0 }
\over W'}} \right].   \eqno (4.6)
$$

\noindent
{\it Proof.}~~~ With $<\phi^i\phi^j> = G^{ij}$ of (3.11), it suffices to show
$$
{\p \over \p t_k} <\phi^i\phi^j> = \mk \left[{{\phi^i\phi^j\phi^k
{\p W\over \p t_0 } \over W'}} \right].
 \eqno (4.7)
$$
Let us start with the case $ i \ge 0,\ j \le -2$.
 By analyitical continuation of the contour in the residue integral we get
$$
\eqalign{
{\rm the\ r.h.s} & = \mi \left[{{\phi^i \phi^j {\p W \over \p t_k} \over W'}}
\right]\ - \
 \ms \left[{{\phi^i \phi^j {\p W \over \p t_k} \over W'}} \right]  \cr
 & {} \cr
& = \mi \left[ {\phi^j{\p \over \p t_k} ({\lambda^{i+1} \over i+1})} \right]\
- \ms \left[{\phi^i {\p \over \p t_k}({\mu^{-j-1} \over j+1})}\right].  \cr}
\eqno (4.8)
$$
The second piece becomes
$$
\ms \left[{\phi^i {\p \over \p t_k}({\mu^{-j-1} \over j+1})} \right]
= \mi \left[{\phi^i {\p \over \p t_k}[{\mu^{-j-1} \over j+1}]_-}\right] =
\mi \left[{{\p \over \p p} ( {\lambda^{i+1} \over i+1})
{\p \over \p t_k}[{\mu^{-j-1} \over j+1}]_-} \right]
$$
Putting this into (4.8) and using (3.11c) yield the l.h.s. of eq. (4.7).
For $j=-1$ we have, with $\phi^{-1} = 1/(p-s)$,
$$
\eqalign{
{\rm the\ r.h.s} & = \mi \left[{{\phi^i{\p W \over \p t_k} \over (p-s)W'}}
\right]\ -\
\ms \left[{{\phi^i{\p W \over \p t_k} \over (p-s)W'}} \right]  \cr
& {} \cr
& = \mi \left[{{ {\p \over\p t_k} ({\lambda^{i+1 }\over i+1} ) \over p-s}}
\right] \ + \ \ms \left[{\phi^i { {\p s \over \p t_k} \over (p-s)}} \right]
 \cr
 & {} \cr
& = {\p \over \p t_k} <\phi^i \phi^{-1}>.  \cr} \eqno (4.9)
$$
For $i=j=-1$ we verify eq. (4.7) by calculating as
$$
\eqalign{
{\rm the\ r.h.s.} & = - \ms \left[{{{\p W \over \p t_k}\over (p-s)^2 W'}}
\right]
 \cr
 & {} \cr
& ={1 \over m}{\p \over \p t_k} \log v_{-m}
= {\p \over \p t_k}<\phi^{-1}\phi^{-1}>.
\cr} \eqno (4.10)
$$
In the second line we have used the formula (3.6a) , i.e. $ v_{-m} = (u_{-m})^m
$.
 For other cases of $i$ and $j$ (4.7) can be shown similarly.
$\quad \square$

\vskip 0.3cm

\noindent
Note here that, if $W$ is a polynomial in $p$, i.e. the $A$-model,
 the residue in eq. (4.6) can be evaluated at $p=\infty$.  We also remark
that $W'$ is nilpotent in the numerator of the integrand in (4.6), and the
 residue formula (4.6) is faithful to the ring structure with the ideal $W'=0$.
With Proposition 4.3, the integrability of the 3-point function is evident,
and is equivalent to (3.14) in Corollary 3.5,
$${\partial  \over {\partial t_i}}<\phi ^j\,\phi ^k\kern 1pt>\;=
\;{\partial  \over {\partial t_j}}<\phi ^i\,\phi ^k\kern 1pt>\;=
\;<\phi ^i\;\phi ^j\,\phi ^k\kern 1pt>\;.  \eqno (4.11)
$$
It should be also noted that the 1-point function  obtained by further
integration of (4.7) can not be explicitly expressed
 without giving a specific form of the solutions of the dKP hierarchy
 (see Section 6).

\medskip

As was shown in the previous section, the coupling constants $t_{\alpha}$ for
$\alpha \in \Delta_{n,m}$ are identified  with the universal coordinates
$u_\alpha$ at a topological limit of TFT (Proposition 3.6).
This suggests that we have another type of free energy
in terms of the universal coordinates $u^\alpha$. Namely we have:

\proclaim Proposition 4.4. There exists a function $F_0 = F_0 (u)$ such
that for $\alpha,\beta \in \Delta_{n,m}$
$$
G^{\alpha\beta} = {\p \over \p u_\alpha}{\p \over \p u_\beta} F_0.
\eqno (4.12)
$$

\noindent
{\it Proof.}~~~ Note first that the generalized GD potential
 $G^{\alpha\beta}$  can be considered as the
functions of the universal coordinates $u^\alpha$ instead of $t_\alpha$.
{}From Proposition 3.2 and the explicit form of $G^{\alpha \beta}$,
one can easily see that $G^{\alpha \beta}$ can be integrated by both
$u^{\alpha}$ and $u^{\beta}$.
$\square$

\vskip 0.3cm

\noindent
Note that the existence of $F_0$ results directly from the definition of
the universal coordinates, and not as a consequence of the dKP hierarchy.
With the more general solution in Proposition 3.6, that is,
${\tilde u}_{\alpha} =<\phi^{\alpha_0}\phi_\alpha>=t_{\alpha}$
for each $\alpha_0 \in \Delta_{n,m}$, we have from Corollary 3.5:
\proclaim Proposition 4.5. There exists a function $F_{\alpha_0}
= F_{\alpha_0} ({\tilde u})$ such that for
$\alpha,\beta \in \Delta_{n,m}$
$$
G^{\alpha\beta} = {\p \over \p {\tilde u}_\alpha}
{\p \over \p {\tilde u}_\beta} F_{\alpha_0}.   \eqno (4.13)
$$

\vskip 0.3cm

\noindent
One should note from these propositions that the 2-point function
$G^{\alpha\beta}$ is a universal object, that is,
$$
G^{\alpha\beta} = {\p^2 F_0 \over \p u_\alpha \p u_\beta} =
{\p^2 F_{\alpha_0} \over \p {\tilde u}_\alpha \p {\tilde u}_\beta} .
\eqno (4.14)
$$
This has been found as a symmetry of the WDVV equation in [Du2].

\medskip

{}From Propositions 4.4 and 4.5, we propose:
\proclaim Definition 4.6. A complex function $<\phi^{\alpha_1} \cdots
\phi^{\alpha_N}>_0$ of the universal coordinates
$u^\alpha$ for $\alpha \in \Delta_{n,m}$ is a universal N-point function
of the fields $\phi^\alpha$, if there exists a function $F_0=F_0(u)$ such that
for $\alpha_i \in \Delta_{n,m}$
$$
<\phi^{\alpha_1}\cdots\cdots\phi^{\alpha_N}>_0(u) =
{\p \over \p u_{\alpha_1}}\cdots \cdots
{\p \over \p u_{\alpha_N}} F_0(u).  \eqno (4.15a)
$$
We call the function $F_0$ the universal free energy. In general, one can
also define universal  N-point functions
for the fields $\{ {\tilde \phi}^{\alpha} \}$ in (2.18) with the free energy
$F_{\alpha_0}({\tilde u})$ as
$$
<{\tilde \phi}^{\alpha_1}\cdots\cdots{\tilde \phi}^{\alpha_N}>_{\alpha_0}
({\tilde u}) =
{\p \over \p {\tilde u}_{\alpha_1}}\cdots \cdots
{\p \over \p {\tilde u}_{\alpha_N}} F_{\alpha_0}({\tilde u}).  \eqno (4.15b)
$$

\vskip 0.3cm

\noindent
 Note that $ F_{\alpha_0}$ gives the free energy $F$ in (4.1) at
the topological  limit
 corresponding to the flat solution ${\tilde u}_\alpha=t_\alpha$.
 For $\alpha_0 =0$, we also notice that owing to (3.3) the 2-point functions
 $G^{\alpha\beta}$ in (3.11) can be
further integrated by $u^{\beta}$ and the universal 1-point functions  are
 expressed as the period integrals,
$$
\eqalignno{
<\phi^\alpha>_0 & = {1 \over (\alpha+n+2)(\alpha+1)} \mi
[ \lambda^{\alpha+n+2}], \quad  0 \le \alpha \le n, & (4.16a) \cr
 & {} \cr
 <\phi^{-1}>_0 & = \mi \left[ {W \left({\log \lambda - {1 \over n+1}}\right)}
 \right] +
 \ms \left[ {W \left({\log \mu - {1 \over m}}\right)} \right] , &  (4.16b)  \cr
 & {} \cr
<\phi^{-\alpha}>_0 & = {1 \over (\alpha+m-1)(\alpha-1)}
\ms[ \mu^{\alpha+m-1}], \quad 2 \le \alpha \le m . &(4.16c) \cr
  & \cr }
 $$
These 1-point functions can be
extended for all fields $\phi^i$ with $i \in \Z$ by integrating $G^{i\alpha}$
with respect to $u_\alpha$.
With these definitions of the N-point functions, we see the universality of
the 2-point functions,
$$
<\phi^\alpha\phi^\beta>_0 = <\phi^\alpha\phi^\beta>
=<{\tilde \phi}^\alpha{\tilde \phi}^\beta>_{\alpha_0}.   \eqno (4.17)
$$
The universal 3-point function can be also written in the residue formula:

\proclaim Proposition 4.7.
$$
<\phi^\alpha\phi^\beta\phi^\gamma>_0
= \mk \left[{{\phi^\alpha\phi^\beta\phi^\gamma
 \over W'}} \right]  \eqno (4.18)
$$

\noindent
{\it Proof.}~~~ Write the r.h.s. as
$$
<\phi^\alpha\phi^\beta\phi^\gamma>_0 = \mk
\left[{{\phi^\alpha\phi^\beta { \p W \over \p u_\gamma} \over W'} } \right] ,
$$
by eq. (3.3). Then eq. (4.18) can be shown similarly as eq. (4.6).
$\quad  \square$

\vskip 0.5cm

\noindent
With this Proposition, the metric defined in (3.4) can be expressed by
the universal 3-point functions, i.e. for $\alpha,\beta \in \Delta_{n,m}$,
$$
\eta_{\alpha\beta} = \eta^{\alpha\beta} = <\phi^{\alpha}
\phi^{\beta}\phi^0>_0.    \eqno (4.19)
$$
Note that in terms of the flat solution $\tilde u_\alpha = t_\alpha$
obtained in Proposition 3.6 we also have
$$
\eta^{\alpha\beta}= {\p \over \p t_\beta} G^{\alpha_0\alpha}
= < \phi^{\alpha} \phi^{\beta} \phi^{\alpha_0} > .   \eqno (4.20)
$$
The flatness of these 3-point functions is a fundamental property for TFT.

\medskip

{}From (4.18), one can define an inner product on the ring $\cal R$ of (2.10)
with a bilinear map, $(\cdot, \cdot)_0 : {\cal R} \times {\cal R} \to {\C}$,
$$
(\phi, \psi)_0 := \mk \left[{\phi\psi \over W'} \right] = <\phi\psi\phi^0>_0.
\eqno (4.21)
$$
The ring with this inner product $\{ {\cal R}, ( \cdot , \cdot )_0 \}$
then defines a commutative
Frobenius algebra.  In particular, the set of primary fields $\{\phi^\alpha\}$
defined in (2.8) gives an orthonormal basis, i.e.
$$
(\phi^{\alpha} ,\phi_{\beta})_0 = \delta^{\alpha}_\beta = \mk \left[
{\phi^{\alpha}\phi_{\beta} \over W' } \right].  \eqno (4.22)
$$
It follows from (4.22) that the following bilinear map also gives an inner
product which makes the fields $\{ {\tilde \phi} \}$ in (2.18)
to be orthonormal, i.e. for a fixed $\alpha_0
\in \Delta_{n,m}$,
$$
({\tilde \phi} , {\tilde \psi})_{\alpha_0} := \mk \left[{ {\tilde \phi}
{\tilde \psi} (\phi^{\alpha_0})^2 \over W'} \right].  \eqno (4.23)
$$
Note that $(\phi, \psi)_0 = ({\tilde \phi}, {\tilde \psi})_{\alpha_0}$,
i.e. the invariance of the inner product under the change of primaries.

\medskip

Let us now consider the fusion algebra (2.21) with the universal
3-point functions (4.18).  Using (4.21), the structure constants
$c^{\alpha\beta}_{\;\gamma}$ in (2.21) can be written by
$$
c^{\alpha\beta}_{\;\gamma} = <\phi^{\alpha}\phi^{\beta}\phi_{\gamma}>_0 =
\sum_{\delta \in \Delta_{n,m}}
<\phi^{\alpha}\phi^{\beta}\phi^\delta>_0
  \eta_{\delta\gamma } .
    \eqno(4.24)
$$
Then the associativity (1.2c)
 follows through reducing the following quantity to
 the universal 3-point functions (4.18),
$$
 \mk \left[{{\phi^{\alpha}\phi^{\beta}\phi^{\gamma}\phi^{\delta}
  \over W'}} \right],  \quad \quad \quad
\alpha, \beta, \gamma, \delta  \in \Delta_{n,m}.   \eqno (4.25) .
$$

\medskip

For the case of the dKP hierarchy in the form (2.17), the fusion algebra
for $\{ {\tilde \phi}^{\alpha} \}$ is defined in the same way:
{}From (4.23), the structure constants ${\tilde c}^{\alpha\beta}_{\;\gamma}$
is given by
$$
{\tilde c}^{\alpha\beta}_{\;\gamma}
= <{\tilde \phi}^{\alpha}{\tilde \phi}^{\beta}{\tilde \phi}_{\gamma}>_{
\alpha_0} =  \mk \left[{ {\tilde \phi}^{\alpha}
{\tilde \phi}^{\beta}{\tilde \phi}_{\gamma}
(\phi^{\alpha_0})^2 \over W'} \right].    \eqno(4.26)
$$
This also gives a residue formula of the universal 3-point functions for
$\{ {\tilde \phi}^{\alpha} \}$, and from (3.17) and (4.17) we have
$$
<{\tilde \phi}^{\alpha}{\tilde \phi}^{\beta}{\tilde \phi}^{\gamma}>_{\alpha_0}
={\p \over \p {\tilde u}_{\gamma}}<{\tilde \phi}^{\alpha}
{\tilde \phi}^{\beta}>_{\alpha_0} = {\p \over \p {\tilde u}_{\gamma}}
< \phi^{\alpha}\phi^{\beta}>_0 \ .  \eqno (4.27)
$$
In particular,  we see $\eta^{\alpha\beta} = <{\tilde \phi}^{\alpha}
{\tilde \phi}^{\beta}{\tilde \phi}^{\alpha_0}>_{\alpha_0}$.

\medskip

 We also note
$$
\phi^{\alpha} \phi^{\beta}  = \sum_{\gamma \in \Delta_{n,m} \ }
c^{\alpha\beta}_{\; \gamma}
\phi^\gamma  = \sum_{\gamma \in \Delta_{n,m} \ }
{\tilde c}^{\alpha\beta}_{\;\gamma}\phi^\gamma \phi^{\alpha_0}
\quad ({\rm mod}\ \ W'),  \eqno (4.28)
$$
which leads to the linear transformation
between the fields $\{\phi^{\alpha} \}$ and $\{ {\tilde \phi}^{\alpha} \}$,
$$
\phi^{\alpha}  =\sum_{\beta \in \Delta_{n,m}} c^{\alpha_0\alpha}_{\;\beta}
{\tilde \phi}^\beta  \quad \quad
( \ {\rm mod} \ W' \ ).  \eqno (4.29)
$$
With (4.29), one can now prove the relation (3.17),
$\p W / \p {\tilde u}_{\alpha}
={\tilde \phi}^{\alpha} \ ({\rm mod } \ W')$:  Taking the derivative of the W
potential with respect to $u_{\alpha}$, we have
$$\eqalign{
\phi^{\alpha} = {\p W \over \p u_{\alpha}}   & = \sum_{\beta \in \Delta_{n,m}}
{\p W \over \p {\tilde u}_{\beta}}{\p {\tilde u}_\beta \over \p u_\alpha} =
\sum_{\beta \in \Delta_{n,m}} {\p W \over \p {\tilde u}_{\beta}}
<\phi^{\alpha_0}\phi^{\alpha}\phi_\beta>_0  \cr
 & {} \cr
 & = \sum_{\beta \in \Delta_{n,m}} {\p W \over \p {\tilde u}^{\beta}}
 c^{\alpha_0\alpha}_{\;\beta} \ \ \ (\ {\rm mod} \ W' \ ). \cr }
 \eqno (4.30)
$$
Using (4.29) then leads to the relation (3.17).

\medskip

Before closing this section, we remark as a corollary to Proposition 4.5 that
the dKP hierarchy in (3.15) for the entire phase space can be put into
a hamiltonian form,
$$
{\p {\tilde u}^\alpha \over \p t_i }= {\p \over \p t_{\alpha_0}}
{\p {\tilde H}^i \over \p {\tilde u}_{\alpha}}
= {\p \over \p t_{\alpha_0}} \sum\limits_{\beta \in \Delta_{n,m}}
\eta^{\alpha\beta} {\p {\tilde H}^i \over \p {\tilde u}^{\beta}},
   \quad i \in {\Z}, \eqno (4.31)
$$
where the hamiltonian function
${\tilde H}^i$ is given by the 1-point function
$<{\tilde \phi}^i>_{\alpha_0}$ defined in (4.15b).  Note here that the
index $i$ in the 1-point function is extended to $ i \in \Z$,
as explained below (4.16).

\vskip 2cm

\noindent
{\bf \S5. The topological recursion relation}

\vskip 0.5cm

In addition to eqs. (1.2), the TFT's also satisfy
the so-called ``topological recursion relation"
which gives a recursion for the 3-point functions of the descendants.
In this section we show that
the dKP hierarchy (2.4) also provides  this important property.

\medskip

Let us first note from  $\p W / \p u_0
= \phi^0 = 1 \ ({\rm mod} \ W')$ that
$$\eqalignno{
\lambda^{i+1} &= \lambda^{i+1} {\p W \over \p u_0} = {\p \over \p u_0}
\left({\lambda^{(n+1)+i+1} \over (n+1)+i+1} \right) ,
\quad {\rm for} \ i \ge 0,  & (5.1a) \cr
    & {}  \cr
 \mu^{j-1} & = \mu^{j-1} {\p W\over \p u_0} = {\p \over \p u_0}
 \left( {\mu^{m+j-1} \over m+j-1} \right) ,  \quad {\rm for} \ j \ge 2.
  & (5.1b) \cr}
 $$
 Setting $ i= N(n+1)+\alpha$ and $ j = Nm + \beta$ with an integer $N \ge 0$
 and $0 \le \alpha \le n, \ 2 \le \beta \le m$, we have the recursion
 relation among $\phi^i$, except $i=-1$,
$$\eqalignno{
\phi^{N(n+1)+\alpha} & = {1 \over N(n+1)+\alpha+1} {\p \over \p u_0}
\phi^{(N+1)(n+1)+\alpha},  & (5.2a) \cr
& {} \cr
\phi^{-Nm-\beta} & = {1 \over Nm+\beta-1} {\p \over \p u_0}
\phi^{-(N+1)m-\beta}. & (5.2b) \cr}
$$
Then we define the fields $\sigma_N(\phi^\alpha)$
for $N \ge 1$, the descendant fields, except $\alpha = -1$,
$$
\sigma_N (\phi^\alpha) := {\p Q_N^{\alpha} \over \p p}
= \left\{ {\matrix{c_{N,\alpha}
\phi^{N(n+1)+\alpha}, \quad\quad  0 \le \alpha \le n , \cr
{}         \cr
d_{N,\alpha}\phi^{-Nm-|\alpha|}, \quad\quad  -m \le \alpha \le -2, \cr} }
\right.  \eqno (5.3)
$$
where the normalization constants $c_{N,\alpha}$ and $d_{N,\alpha}$ are
defined by
$$
\eqalignno{
c_{N,\alpha} & =
[(\alpha+1)(\alpha+1+n+1)\cdots\cdots(\alpha+1+(N-1)(n+1))]^{-1}, & (5.4a) \cr
d_{N,\alpha} & =
[(|\alpha|-1)(|\alpha|-1+m)\cdots\cdots(|\alpha|-1+(N-1)m)]^{-1}.  & (5.4b)
 \cr}
$$
For the case of $\alpha = -1$, the descendant field can be also defined
in the same way. Namely we have, modulo $W'$,
$$\eqalignno{
W^N (\log \lambda - c_{N}) &= {1 \over N+1} {\p \over \p u_0}
 \left( W^{N+1}
( \log \lambda - c_{N+1}) \right) , & (5.5a) \cr
& {} \cr
W^N (\log \mu - d_{N}) &= {1 \over N+1} {\p \over \p u_0} \left( W^{N+1}
( \log \mu - d_{N+1}) \right), & (5.5b) \cr }
$$
where the constants $c_{N}$ and $d_{N}$ are defined by
$$\eqalignno{
c_{N} & = {1 \over n+1} \sum_{l=1}^N {1 \over l}, & (5.6a) \cr
d_{N} & = {1 \over m} \sum_{l=1}^N {1 \over l} \ , & (5.6b) \cr }
$$
for $N \ge 1 $ and $c_0 = d_0 = 0$.
The descendant fields $\sigma_N(\phi^{-1})$ are then defined by
$$\eqalign{
\sigma_N(\phi^{-1}) :&= {\p Q_N^{-1} \over \p p}  \cr
&= {\p \over \p p} \left( \left[{W^N \over N!} (\log \lambda - c_{N}) \right]_+
 - \left[{W^N \over N!}(\log \mu - d_{N}) \right]_- \right) . \cr } \eqno (5.7)
$$
In Appendix A, we give the explicit formula of (5.7), and show that
$\sigma_N(\phi^{-1})$ is defined as an element of the ring $\cal R$ of (2.10).
The compatibility of this field with others can be also shown in the similar
 way as in Lemma 2.1. Eq.(5.7) gives a precise definition and generalization
of the descendant fields of $\phi^{-1}$ found in [EY].
With these definitions (5.3) and (5.7), eqs.(5.2) and (5.5)
lead to a recursion relation,
$$
\sigma_{N-1}(\phi^\alpha) = {\p \over \p u_0} \sigma_N (\phi^\alpha),
\quad ( \ {\rm mod} \ W' \ ) . \eqno (5.8)
$$
Correspondingly to those descendant fields, we also define
the gravitational coupling constants $t_{N,\alpha}$  as
$$t_{N,\alpha }\;=\;\left\{ {\;\matrix{{\;c_{N,\alpha }^{-1}\,
t_{N(n+1)+\alpha }\kern 1pt,\quad0\;\le \kern 1pt\;\alpha \;\le \;n\;,}\cr
{} \cr
{\matrix{{\matrix{{\matrix{ { \quad \quad t_{N,-1}, }
&{}&{ \quad \quad \quad \quad \alpha = -1 }\cr
}}&{}&{}\cr
}}&{}&{}\cr
}}\cr
{} \cr
{d_{N,\alpha }^{-1}\,t_{-Nm-|\alpha|}\kern 1pt,\quad-m\;\le
\kern 1pt\;\alpha \;\le \;-2\;,}\cr
}} \right.  \eqno (5.9)
$$
Thus the gravitational descendants are constructed entirely from the primary
(matter)
fields alone, similar to the case of the minimal model [Lo, EKYY, EYY1].

\medskip

With the definitions of $Q^{-1}_N$ in (5.7), and $t_{N,-1}$ in (5.9), one can
extend the 3-point function in Proposition 4.3 to include the field
 $\sigma_N(\phi^{-1})$.  In particular, following the calculations in
Appendix A and in the proof of Proposition 4.3, we find the formula for
the 2-point function with $\sigma_N(\phi^{-1}) $,
$$
<\sigma_N(\phi^{-1})\phi^i> =  \mi \left[{{W^N \over N!} ({\log \lambda -
c_{N} })\phi^i}\right] +
\ms \left[{{W^N \over N!} ({\log \mu - d_{N} })\phi^i
}\right] .  \eqno (5.10)
$$
As shown in (4.14), the 2-point function $<\sigma_N(\phi^{-1})\phi^{\alpha}>$
can be further integrated by $u_\alpha$, and
 the flow $\p u^{\alpha} / \p t_{N,-1}$ is expressed in the hamiltonian form of
(4.34) with $H_N^{-1}$,
$$\eqalign{
 & H_N^{-1} = <\sigma_{N}(\phi^{-1})>_0  \cr
  & {}  \cr
& = \mi \left[{{W^{N+1} \over (N+1)!}
 ( \log \lambda - c_{N+1} )}\right] +
\ms \left[{{W^{N+1} \over (N+1)!} ( \log \mu - d_{N+1} )
}\right]. \cr } \eqno (5.11)
$$

\medskip

 By the ideal $W'=0$,
we have the extended relation of (2.9),
$$
\eqalign{
\sigma_N(\phi^{n}) & = \sigma_N (\phi^{-(m+1)}) \quad ({\rm mod}\ W') ,
\quad \quad   N \ge 0,  \cr}  \eqno (5.12)
$$
with the identifications  $\sigma_0(\phi^\alpha)=\phi^\alpha$, $c_{0,\alpha}
=d_{0,\alpha}=1$ for any $\alpha \in \Delta_{n,m}$, and
$\sigma_N(\phi^{-(m+1)}) = \phi^{-(N+1)m-1} / N!m^N$. This implies that
  the solutions of the dKP hierarchy (2.14) have the form with
$t_{0,\alpha}:=t_\alpha$,
$$
u^\alpha= u^\alpha (\ t_{N,\beta} :\ t_{N,n} + t_{N,-(m+1)},
\ {\rm for \  each} \ N \ge 0, \ {\rm and} \ \beta \in \Delta_{n,m}
\ ).  \eqno (5.13)
$$
where $t_{N,-(m+1)}$ is the flow parameter corresponding to the field
$\sigma_N(\phi^{-(m+1)})$, i.e. $t_{N,-(m+1)} = N!m^N t_{-(N+1)m-1}$.
Note that these $\sigma_N(\phi^{-(m+1)})$, and $t_{N,-(m+1)}$ were
readily excluded from the defining relations (5.3) and (5.9).
However, because of (5.13)
we identify $t_{N,n}$ with $t_{N,n} + t_{N,-(m+1)}$.
 With these notations  one obtains the main theorem of this section:
\proclaim Theorem 5.1. For each primary $\phi^\alpha$, we
have the topological recursion relation for the 3-point functions,
$$
<\sigma_N (\phi^\alpha) AB> = \sum_{\beta \in \Delta_{n,m}}
<\sigma_{N-1}
 (\phi^\alpha) \phi^\beta> <\phi_{\beta}AB> ,  \eqno (5.14)
$$
for any $A, B \in {\rm \C}[p,(p-s)^{-1}]$ and $N \ge 1$.

\noindent
To prove this theorem we need:
\proclaim Lemma 5.2. The descendants can be decomposed into
the primaries, i.e.,
$$
\sigma_N (\phi^\alpha) = \sum_{\beta \in \Delta_{n,m}}
<\sigma_{N-1} (\phi^\alpha)\phi^\beta> \phi_\beta
 \quad \quad  ({\rm mod}\ W'),   \eqno (5.15)
$$

\noindent
{\it Proof.}~~~
{}From the orthonormality (4.22), we note that eq. (5.15) is equivalent to
$$
\eqalignno{
<\sigma_{N-1} (\phi^\alpha)\phi^\beta>
 & = <\sigma_N(\phi^\alpha)\phi^\beta\phi^0>_0  \cr
  & {} \cr
& =
\mk \left[{{\sigma_N (\phi^\alpha)\phi^\beta \over W'}} \right].
 & (5.16) \cr }
$$
The function in the residue has a pole at either $p=\infty$ or
$p=s$ in addition to those at the
roots of $W'= 0$. Therefore we evaluate (5.16) for the case of $\alpha \ne -1$
$$
{\rm the\ residue} = \left\{ {\matrix{
{c_{N,\alpha}\mi
[\lambda^{(N-1)(n+1) + \alpha +1} \phi^\beta] , \quad \quad
{\rm for}\quad \alpha \ge 0}
\cr
{}  \cr
 {d_{N,\alpha}\ms
[\mu^{(N-1)m - \alpha -1} \phi^{\beta}], \quad \quad
{\rm for}\quad  \alpha \le -2}
 \cr}}\right.
 $$
which gives, with (3.11), the l.h.s. of (5.16).
For the case of $\alpha=-1$, we have
$$\eqalign{
&<\sigma_N(\phi^{-1})\phi^{\beta}\phi^0 >_0 = \mk
\left[{{\sigma_N(\phi^{-1})\phi^{\beta}\phi^0 \over W'}}\right] \cr
 & {} \cr
& = \mi
\left[{{W^{N-1} \over (N-1)!} (\log \lambda - c_{N-1}) \phi^{\beta}}\right] +
\ms
\left[{{W^{N-1} \over (N-1)!} (\log \mu - d_{N-1}) \phi^{\beta}}\right] \cr
& {} \cr
&= <\sigma_{N-1}(\phi^{-1})\phi^{\beta}>.  \quad \quad \square  \cr }
$$

\medskip

\noindent
Then the proof of Theorem 5.1 is straightforward with the residue
formula of the three point functions (4.6).
{}From (5.16),  we also note the recursion relation for the hamiltonian
functions $ \ H_N^{\alpha} := \ <\sigma_N(\phi^{\alpha})>_0$ for all $\alpha
\in \Delta_{n,m}$ and $N \ge 1$,
$$
{\p H_N^{\alpha} \over \p u_0 } = H_{N-1}^{\alpha}.    \eqno (5.17)
$$

\vskip 2cm

\noindent
{\bf \S6.~ The string equations}

\vskip 0.5cm

Here we derive the ``string equation" as the solution of
the dKP hierarchy (3.10),
and give an explicit scheme to construct the corresponding free energy.
The main result in this section is to show that all the effects of the
gravitational couplings to the constitutive equations (2-point
functions) can be described in the small phase space alone
by renormalizing the primary couplings ,
that is, the renormalizability of the solutions of our TFT.
Let us first note from Theorem 5.1:

\proclaim Corollary 6.1. The dKP hierarchy (3.10) for the gravitational
couplings  $t_{N,\beta}$ can be decomposed into the flows for the primaries
with $t_\gamma$,
$$
{\p u^\alpha \over \p t_{N,\beta}} = \sum_{\gamma \in \Delta_{n,m}}
<\sigma_{N-1}(\phi^\beta)\phi_\gamma> {\p u^\alpha \over  \p t_\gamma},
\quad \quad
\alpha, \beta \in \Delta_{n,m},  \eqno (6.1)
$$
where the 2-point funtion $<\sigma_{N-1}(\phi^\beta)\phi_\gamma>$
is a function of $u^\alpha$ given by (4.3).

\noindent
{}From Corollary 6.1 one can obtain:

\proclaim Theorem 6.2. [KG]
The solution of the dKP hierarchy can be expressed by
$$
\eqalignno{
u^\alpha (t_j: j \in \Z \ ) & = \hat u^\alpha
(\hat t_\beta:\beta \in \Delta_{n,m}),
\quad {\rm for \ }\alpha \in \Delta_{n,m}, \quad  & (6.2)  \cr}
$$
where $\hat u^\alpha$ and $\hat t^\beta$ are given by for all
$\alpha,\beta \in \Delta_{n,m}$
$$
\eqalignno{
& \hat u^\alpha (t_\gamma: \gamma \in \Delta_{n,m} \ )  = u^\alpha
( \ \cdots, 0,t_{-m}, \cdots, t_n, 0, \cdots \ )
 & (6.3a)  \cr
 & {} \cr
& \hat t_\beta  =  t_\beta + \sum_{\alpha \in \Delta_{n,m} \atop N \ge 1}
 <\sigma_{N-1}(\phi^\alpha)\phi_\beta>
t_{N,\alpha}  & (6.3b)     \cr}
$$

\noindent
{\it Proof.}~~~ Eq. (6.1) can be expressed in the invariant form of
the vector field $X_N^\beta$,
$$
X_N^\beta u^\alpha = 0  \quad \quad \quad {\rm for}\ \alpha, \beta \in
 \Delta_{n,m} \ ,  {\rm and} \ N \ge 1,\eqno (6.4)
$$
with
$$X_N^\beta :={\partial  \over {\partial t_{N,\beta }}}\
-\ \sum\limits_{\gamma \in \Delta_{n,m}}
{<\sigma _{N-1}(\phi ^\beta )\phi_\gamma >{\partial  \over
{\partial t_\gamma }}}  \eqno (6.5)
$$
This implies that each $u^\alpha$ is constant along the characteristic,
which are straight lines,
given by, for $\alpha,\beta \in \Delta_{n,m}$ and $ N\ge 1$,
$$
{d t_{N,\beta} \over -1} = { d t_\alpha \over <\sigma_{N-1}(\phi^\beta)
\phi_\alpha>} \ .  \eqno (6.6)
$$
The integrals of eq. (6.6) are
$$
\hat t_\alpha = t_\alpha +
 \sum_{\beta \in \Delta_{n,m}} <\sigma_{N-1}(\phi^\beta)\phi_
 \alpha>t_{N,\beta},  \eqno (6.7)
$$
which gives eq. (6.3b) on taking sum over $N \ge 1$ .
Here $\hat t_\alpha$ gives
the initial position of the characteristics at
$t_{N,\beta}= 0$ for all $N \ge 1$. Note then that
 $\hat u^\alpha$ in (6.3a) are the solutions of
dKP hierarchy in the small phase space labeled
 by $\{\hat t_\alpha : \alpha \in \Delta_{n,m} \}$. This completes the proof.
$\quad \square$

\vskip 0.5cm

\noindent
This theorem implies that the solution of the dKP hierarchy is completely
 determined by that in the small phase space, $\hat u^\alpha$.
To be concrete, we first note that
 eq. (6.3b) may be written in the following form, which gives the string
equation
 of our TFT (see the end of this section),

$$
0 = \tilde t_\alpha +
 \sum_{\beta \in \Delta_{n,m} \atop N \ge 1} <\sigma_{N-1}(\phi^\beta)\phi_
 \alpha>\tilde t_{N,\beta},  \eqno (6.8)
$$
by shifting the couplings with arbitrary constants $C_{N,\beta}$ as
$$
\tilde t_{N,\beta} = t_{N,\beta} + C_{N,\beta}, \quad \quad
{\rm for}\quad N \ge 1, \eqno (6.9a)
$$
and by imposing the relations,
$$
-\sum_{\beta \in \Delta_{n,m} \atop N \ge 1}
 <\sigma_{N-1}(\phi^\beta)\phi_\alpha> C_{N,\beta}
= \hat t_\alpha.  \eqno (6.9b)
$$
Note here that the dKP hierarchy is translationally invariant in the couplings,
that is, the solution can be written in the shifted variables
$\tilde t_{N,\alpha}$. A solution $\hat u^\alpha (\hat t_\gamma)$
of the dKP hierarchy in the small phase space
is then given by solving the algebraic equations (6.9b).
For example, by the choice of $C_{N,\beta} =
 -\delta_{N,1}\delta_{\beta,\alpha_0}$, this equation coincides
with (3.18), so that the dKP hierarchy provides a flat solution of our TFT.
All the flat solutions are indeed obtained from this choice of the constants
$C_{N,\beta}$.  The solution in the large phase space
is given by merely writing $\hat t_\gamma$ in $\hat u^\alpha (\hat t_\gamma)$
by (6.3b) . This is an implicit solution called the hodogragh solution,
 which  still includes the function $\hat u^\alpha$ in the r.h.s.
of (6.3b). (See below for the construction of explicit solutions.)
Thus depending on the values of $C_{N,\beta}$ one can construct infinitely
many solutions of the dKP hierarchy in the entire phase space.
 Physically speaking, a choice of $C_{N,\beta}$ amounts to considering a TFT
 in the small phase space where the gravitational couplings take the
 fixed values,
 $$
 \tilde t_{N,\beta} = C_{N,\beta}, \quad \quad   N \ge 1,  \quad
 \beta \in \Delta_{n,m} , \eqno (6.10)
 $$
 and the TFT in the large phase space is obtained as a perturbation
 from this gravity background.

 \medskip

 The physical meaning of (6.2) is that all the gravitational effects
in the universal coordinates $u^{\alpha}(t)$, consequently
  in the 2-point function $<\phi^\alpha \phi^\beta >$  are renormalized
  into the primary couplings $\hat t_{\alpha}$  by (6.3b).
This renormalizability  of the TFT can be most properly seen
by writing the action (2.13) as
$$
S = S_0 + \sum_{\alpha \in \Delta_{n,m} \atop N \ge 0 }
t_{N,\alpha}\sigma_N(\phi^\alpha)
  =  S_0 +  \sum_{\alpha \in \Delta_{n,m}}\hat t_\alpha \phi^\alpha \quad
  ({\rm mod}\  W').  \eqno (6.11)
$$
This follws from Lemma 5.2 for the  decomposition of the descendants. By
(6.3b) and (6.9b) we obtain the string equation in the generalized form,
$$
-\sum_{\beta \in \Delta_{n,m} \atop N \ge 1}
 <\sigma_{N-1}(\phi^\beta)\phi_\alpha> C_{N,\beta}
 =  t_\alpha + \sum_{\beta \in \Delta_{n,m} \atop N \ge 1}
 <\sigma_{N-1}(\phi^\beta)\phi_\alpha> t_{N,\beta}.  \eqno (6.12)
$$

\vskip 0.4cm

\noindent
Here the 2-point functions  are the known function of $u_\alpha$ by
the explicit forms (3.11) and
(5.10), thereby (6.12) gives an implicit solution of the dKP hierarchy,
the generalized hodograph solution.
In order to  solve (6.12) explicitly, we employ a perturbation method,
assuming the gravitational couplings $t_{N,\alpha}$ to be small,
where the leading solution is given by (6.9b) with $\hat t_\alpha = t_\alpha$,
i.e. $t_{N,\alpha}=0$ for $N \ge 1$.
 Thus  one obtains  the solution $u^\alpha (t_i)$
 as a formal series in the gravitational couplings $t_{N,\alpha}$.
 This is the well-known renormalization procedure  in the quantum field theory.
   For example, in the  case
   where $ C_{N,\beta} = - \delta_{N,1}\delta_{\beta,0}$,
  eq. (6.9b) gives the simplest solution
$$
\hat u_\alpha = \hat t_\alpha.
$$
The string equation (6.12) then becomes
$$
 u_\alpha  =  t_\alpha + \sum_{\beta \in \Delta_{n,m} \atop N \ge 1}
 <\sigma_{N-1}(\phi^\beta)\phi_\alpha> t_{N,\beta},  \eqno (6.13)
$$
which was discussed in [DW, EYY2]. The above mentioned renormalization can
be easily carried out in this case.

\medskip

Let us now  give an explicit form of the free energy $F$ of our TFT.
Based on the string equation (6.12) we have:
\proclaim  Theorem 6.2. [Du1, Kr1, TT2]
The free energy corresponding to eq. (6.9b) is given by
$$
\eqalign{
& F(t_{N,\alpha} : \alpha \in   \Delta_{n,m}, N \ge 0)   \cr
& {} \cr
& = {1 \over 2} \sum_{ \alpha,\beta  \in \Delta_{n,m}, \atop N,M \ge 0}
\tilde t_{N,\alpha} \tilde t_{M,\beta}
<\sigma_N(\phi^\alpha)\sigma_M(\phi^\beta)>(t), \cr}  \eqno (6.14)
$$
where $\tilde t_{0,\alpha} = t_{\alpha}$.

\noindent
{\it Proof.}~~~ By multiplying (6.8) by $<\phi^\alpha AB >$ with any fields
$A, B $
 and using the topological recursion relation (5.15), the string
 equation (6.8) becomes
$$
0 = \sum_{\beta \in \Delta_{n,m} \atop N \ge 0} \tilde t_{N,\beta}
<\sigma_N (\phi^\beta)AB> =\sum_{\beta \in \Delta_{n,m} \atop N \ge 0}
\tilde t_{N,\beta}{\p \over \p t_{N,\beta}} <AB>. \eqno (6.15)
$$
This shows that the 2-point function is a homogenious function of
degree zero in $t_{N,\beta}$, so that the free energy is of degree two,
that is, with an Euler operator ${\cal E}[\cdot]$,
$$
{\cal E}[ F ]:= \sum_{\beta \in \Delta_{n,m} \atop N \ge 0}
\tilde t_{N,\beta}{\p \over \p t_{N,\beta}} F = 2 F .  \eqno (6.16)
$$
The formula (6.14) is then obtained by applying the Euler
operator once again to (6.16).  $ \quad \square$

\vskip 0.5 cm

\noindent
One should again note that the 2-point functions $<\phi^i\phi^j>$
in eq. (6.14) are the explicitly given quantities in terms of the
universal coordinates $\{ u^\alpha\}$, which are the solutions of the dKP
hierarchy.  We also note that  eq. (6.14) can be reduced to the free energy
 on the small phase space with $t_{N,\alpha} = 0$ for $N \ge 1$,
$$
\eqalign{
\hat F (t_\gamma:\gamma \in \Delta_{n,m}) & = {1\over 2} \sum_{\alpha,\beta
\in \Delta_{n,m}}
 t_\alpha t_\beta<\phi^\alpha\phi^\beta>
 + \sum_{\alpha,\beta \in \Delta_{n,m} \atop N \ge 1} C_{N,\alpha}t_\beta
 <\sigma_N (\phi^\alpha) \phi^\beta>  \cr
& + {1 \over 2} \sum_{\alpha,\beta \in \Delta_{n,m} \atop N,M \ge 1}
C_{N,\alpha}C_{M,\beta}<\sigma_N (\phi^\alpha)\sigma_M (\phi^\beta)>.
 \cr } \eqno (6.17)
$$
In the section 8, we discuss several examples of LG potentials,
 and give the explicit formulae of the free energy.

\medskip

As a final remark we note that
the string equation (6.8) can be put in the form,
$$
L_{-1} Z := \left( \sum_{\beta \in \Delta_{n,m} \atop N \ge 1}{\tilde t}_
{N,\beta}{\p \over \p t_{N-1,\beta}}  +  {1 \over 2}\sum_{\alpha,
\beta \in \Delta_{n,m}\ }
\eta^{\alpha\beta}{\tilde t}_\alpha {\tilde t}_\beta \right) Z = 0 ,
\eqno (6.18)
$$
where $Z$ is the partition function defined by $F = \log Z$.  This would be
generalized to
the Virasoro constraints $L_i Z =0$ for $i \ge -1$.  However there is a subtle
problem to determine the Virasoro operators with $i \ge 0$ in a TFT having a
scale-violation like some of our TFT's.   In the recent paper [EHY], this was
discussed for the $CP^{1}$-model, and the explicit forms of the Virasoro
operators were obtained.  It is interesting to investigate this issue for our
general model.

\vskip 2cm

\noindent
{\bf \S7. Critical phenomena}

\vskip 0.5cm

In this section we discuss the critical behavior of the TFT coupled to
the gravity based on the string equation (6.12). This corresponds to
studying that of a matrix model in the genus-zero (classical) limit,
which would obey a constrained KP hierarchy with the
W potential in a pseud-differential form.

\medskip

We call our TFT with the rational potential (2.1) as $W_{n,m}$-model.
 For example, $W_{n,0}$ gives the $A_n$-model, and
$W_{2n+1,2}$ with the ${\Z}_2$-symmetry, where the deformation variables
are constrained
by $s = 0 $ and $v_{2\alpha+1 } = 0$ for $-1 \le \alpha \le n-1$, gives
the $D_n$-model. Also, $W_{2n+1,2m}$ with the ${\Z}_2$-symmetry is a natural
 extension of the latter [T]. ( The truncation by the ${\Z}_a$-symmetry with
 $a \ge 3$ does not make sense, since the flat metric is
 vanishing for the primaries given by $\phi^\alpha$ , where $\alpha$ is
 the $a$-multiple.
In this regard there is a misstatement in our previous paper [AK].)
As is clear from the free energy (6.14), the $W_{n,m}$-model with non-zero $m$
in general has a scaling violation due to the log-term.
The above models are all of the types which  do not have such a violation
in our TFT.

\medskip

We now study  critical phenomena of the $W_{n,m}$-model by scaling to
the special gravity background given by (6.10),
$$
\tilde t_{N,\beta} = - \delta_{N,N_0}\delta_{\beta ,\beta_0}
\quad {\rm for \ some} \ \beta_0, \ N_0 \ge 1.
$$
This amounts to studying the dKP hierarchy in the small phase space with
$C_{N,\beta} =  - \delta_{N,N_0}\delta_{\beta ,\beta_0}$.
The solution of the hierarchy is then given by the string equation (6.12),
i.e.,
$$
<\sigma_{N_0 -1}(\phi^{\beta_0})\phi_\alpha > = t_\alpha.  \eqno (7.1)
$$
For the scaling models, we find the critical behavior of the free energy,
$$
F    \longrightarrow  \epsilon^{2-\gamma_{\rm string}} F,  \quad {\rm as} \quad
 t_\alpha  \longrightarrow \epsilon^{1-\gamma_\alpha} t_\alpha \ . \eqno (7.2)
$$
Here $\epsilon$ is a (dimensionless) scaling
parameter, $\gamma_{\rm string}$ is called the string susceptibility [FGZ],
 and $\gamma_{\alpha}$ is the critical exponent of the primary $\tilde
\phi^{\alpha}$
 following from (2.17), i.e.
$$
\tilde \phi^\alpha \longrightarrow  \epsilon^{\gamma_\alpha}
\tilde \phi^\alpha.  \eqno (7.3)
$$
Since $\tilde \phi^{\alpha_0} = {\cal P} \ (=1)$ the puncture operator,
we take  $\gamma_{\alpha_0}=0$.
Hence the dimension of $t_{\alpha_0}$ equals to $1$, and
 $t_{\alpha_0} = t_{\cal P}$, the
cosmological constant.
The values of $\gamma_{\rm string}$ and $\gamma_\alpha$ can be
computed by a dimensional analysis of the string equation (7.1).
(Make use of the case  $\alpha = \alpha_0$ to fix the dimention of
$\lambda$ and $\mu$.)   Here we give only the results:

\medskip

a) Case with $\alpha_0 \ge 0$:
$$
\eqalign{
\gamma_{\rm string} & = -{2 \over d_{+\pm }} \ ,  \cr
\gamma_\alpha   & = { \alpha - \alpha_0 \over d_{+\pm} } \ ,
 \quad \quad \quad\quad \quad\quad \quad \quad\quad\quad \quad \quad
{\rm for} \quad \alpha \ge 0 \ , \cr
\gamma_{\alpha} &  = -{ (\alpha + 1) {n+1 \over m} + \alpha_0
+ 1 \over d_{+\pm} } \ ,
\quad \quad \quad \quad\quad  {\rm for} \quad \alpha < 0 \ ,  \cr }  \eqno
(7.4)
$$
where
$$
\eqalign{
  d_{++} & = N_0 (n+1) - \alpha_0 + \beta_0 \ , \quad \quad \quad \quad
\quad \quad\quad\quad  {\rm for} \quad \beta_0 \ge 0 \ , \cr
 d_{+-} & = N_0 (n+1) - \alpha_0 - 1 - (\beta_0 + 1){n+1 \over m} \ ,
 \quad \quad {\rm for} \quad \beta_0 < 0  \ . \cr}
$$

\medskip

b) Case with $\alpha_0 < 0$:
$$
\eqalign{
\gamma_{\rm string} & = -{2 \over d_{-\pm }} \ ,   \cr
\gamma_{\alpha} &  = { (\alpha + 1) {m \over n+1}
+ \alpha_0 + 1 \over d_{-\pm} } \ ,
\quad \quad \quad \quad\quad  {\rm for} \quad \alpha \ge 0 \ ,  \cr
\gamma_\alpha   & = -{ \alpha - \alpha_0 \over d_{-\pm} } \ ,
\quad \quad \quad\quad \quad\quad \quad \quad\quad\quad \quad \quad
{\rm for} \quad \alpha < 0 \ , \cr} \eqno (7.5)
$$
where
$$
\eqalign{
 d_{-+} & = N_0 m + \alpha_0 + 1 + (\beta_0 + 1){m \over n+1} \ ,
 \quad \quad {\rm for} \quad \beta_0 \ge 0  \ ,  \cr
  d_{--} & = N_0 m + \alpha_0 - \beta_0 \ ,
  \quad \quad \quad \quad \quad \quad\quad\quad
{\rm for} \quad \beta_0 < 0  \ . \cr}
$$

\medskip

For the $W_{n,0} (A_n)$-model these results are well known in the critical
analysis of matrix models [Do, GGPZ, FK, FKN, DVV2].  In fact
the critical $W_{n,0}$-model is obtained as the genus-zero limit of
the double scaled $(p-1)$-matrix
model at the $q$th criticality in which
$$
p-1 = n, \quad \quad \quad  q-1 = (N_0 -1)(n+1) + \beta_0.  \eqno (7.6)
$$
The latter model is identified with the $(p, q)$ minimal model
coupled to the gravity. Among the minimal ones the $(n+1, n+2)$ model,
for $N_0 = 2$ and $\beta_0 = 0$, falls into the unitary series,
of which the central charge is given by $1 - { 6 \over (n+1)(n+2)} $
in the gravitationless limit [KPZ, DK, Da].  The $(n+1, q)$ models with
$ 1 \le q \le n+1$ correspond to the topological limits of the $W_{n,0}$-model.

\medskip

For the $W_{2\nu +1,2\mu}$-model with the ${\Z}_2$-symmetry these results
should be understood with the following parametrization of the indices,
$$\eqalign{
n & =2\nu + 1, \quad m = 2\mu, \quad {\rm and} , \cr
\alpha & = 2a , \quad {\rm for} \    a \in \Delta_{\nu,\mu} \ . \cr} \eqno
(7.7)
$$
An interesting observation about the models of this type is that there is a
symmetry in the critical exponents given by (7.4) and (7.5) under the
simultaneous interchange
$$\eqalign{
& n \to m -1, \quad  m \to n +1,  \quad {\rm and} \cr
& \alpha + 1   \to - (\alpha + 1 ) \ .
\cr } \eqno (7.8)
$$
This implies that the $W_{2\nu +1, 2\mu}$-model in the gravity background given
by $\tilde t_{N,\beta} = - \delta_{N,N_0}\delta_{\beta,\beta_0}$ has the same
critical behavior as the $W_{2\mu -1,2\nu +2}$-model in the background
$\tilde t_{N,\beta} = - \delta_{N,N_0}\delta_{\beta,-(\beta_0+2)}$, if the
primary coupling $t_{\alpha}, \alpha \in \Delta_{2\nu,2\mu}$ in the former is
identified with $t_{-(\alpha +2)}, -(\alpha +2) \in \Delta_{2\mu -2, 2\nu +2}$
in the latter.  This symmetry holds for any $N_0 (\ge1)$.
Hence both models are considered to be physically equivalent even when
the gravity is
turned on, by generalizing the identification of the primary couplings
for the descendant couplings.  Thus there exists an equivalent pair
in the $W_{n,m}$-models with the ${\Z}_2$-symmetry.

\medskip

Although the scaling analysis from  (7.1) to (7.5) cannot be applied for
the case without the ${\Z}_2$-symmetry, we conjecture that in general the
$W_{n,m}$-model is equivalent to the $W_{m-1,n+1}$-model by the above
identification of the couplings.  Notice that this equivalence is based on
the interchange of the local coordinates $\lambda$ and $\mu$ in (2.2) (i.e.
interchange between $p$ and $p-s$).  It is also interesting to see how
this symmetry is extended for the LG models with multi-poles in
the rational potential [Kr2].  In the next section, this equivalence will be
checked for the $W_{0,2}$- and the $W_{1,1}$-models, by calculating the
free energy in the small phase space with  the different gravity
backgrounds given by
$\tilde t_{N,\beta} = - \delta_{N,1}\delta_{\beta,\alpha_0}$.

\medskip

We now discuss  a formation of singularity in the solution of the dKP
hierarchy.
As an equation of quasi-linear hyperbolic system,
the solution of the dKP hierarchy
in general breaks in finite time (formation of shocks).  This singularity may
represent a phase transition of the matter states due to the gravitational
couplings. To regularize this singularity, one needs to include an
effect of finite genus, that is, a quantum correction to the
phase transition, which can be studied
by extending the dKP hierarchy to the Whitham hierarchy [BKo, Du1, Kr2].
The genus in this case would concern with the target space of
the TFT, instead of the world sheet.
In the following, we illustrate this phase
transition (shock formation) for the  $W_{1,0}$-model (pure gravity).

\medskip

In this model, the ring ${\cal R}$ consists of only one primary field, that is,
the puncture operator $\phi^0 = \phi_0 = {\cal P}$.  Then the string equation
(6.8) gives, with $<\sigma_{N-1}(\phi^0)\phi_0> = u_0^N/N!$,
$$
0 = {\tilde t_0} + \sum\limits_{N \ge 1} {\tilde t}_{N,0}{u_0^N \over N!},
\eqno (7.9)
$$
where ${\tilde t}_{N,0} = t_{N,0} + C_{N,0}$.  For example, if we take $C_{N,0}
= - \delta_{N,N_0} N_0!$ with a fixed number $N_0 > 0$, (7.9) expresses
the $N_0$th critical phenomena,
$$
u_0^{N_0} = t_0 + \sum\limits_{N \ge 1} {u_0^N \over N!} t_{N,0}.   \eqno
(7.10)
$$
To discuss a deformation of the critical phenomena based on this equation, let
us choose $N_0=3$ and set all parameters $t_{N,0} = 0$ except at
$N= 0$ and $1$, i.e. with $t_{0,0} := t_0$ and $t_{1,0} :=  t_1$
$$
u_0^3 = t_0 + u_0 t_1.   \eqno (7.11)
$$
Fig. 1 shows the bifurcation diagram obtained from (7.11)
in the phase space $(t_1, t_0)$.

\vskip 8cm

\centerline { Fig.1 : The bifurcation diagram for the $W_{1,0}$-model}
\centerline{ with the $N_0 = 3$ criticality.}

\vskip 1cm

\noindent
The curve in the figure gives the points where the derivative $\p u_0 / \p t_0$
becomes infinity.  This indicates that the criticality at $t_0 = 0$
can be resolved by taking the gravitational coupling to be $t_1 < 0$, and
when $t_1 > 0$ it bifurcates into two regions bounded by the singular curve.
The region including the $t_0$-axis corresponds to the
regular state of the matter field, while the region including
the positive $t_1$-axis gives a different state which may imply a genus
creation in the target space.  The free energy of this system is then
caluculated as,
$$
F={4 \over 105} t_0 t_1^2 + \left({{9 \over 28} t_0^2 + {4 \over 105}
t_1^3}\right) u_0 + {27 \over 140} t_0 t_1 u_0^2,  \eqno (7.12)
$$
where $u_0$ is given by the solution of (7.11).  In a future communication,
we will discuss a process of this bifurcation.

\vskip 2cm

\noindent
{\bf \S8. Examples}

\vskip 0.5cm

In order to demonstrate the results obtained in this paper,
 we here give the explicit results for several models,
 which include the $W_{0,1}$-, $W_{2,0}$-models for the examples
with two primaries, and the $W_{0,2}$-, $W_{1,1}$-models for those
with three primaries.

\vskip 0.3cm

\noindent
{\it $W_{0,1}$-model} :

The W potential in this case is given by (2.1),
$$
W=p+{v_{-1} \over {p-s}}, \quad \quad  \eqno (8.1)
$$
where $v_{-1}$ and $s$ are related to the universal coordinates as (3.6),
$$
v_{-1}= u^0 = u_{-1}, \quad  s=u^{-1} = u_0.    \eqno (8.2)
$$
Then we have (2.2),
$$\eqalignno{
\lambda & = p + {u_{-1} \over p} + {u_{-1}s \over p^2} + \cdots, & (8.3a) \cr
\mu & = {u_{-1} \over {p-u_0}} + u_0 + (p-u_0).    & (8.3b) \cr }
$$
The flow equation for $t_{-1}$ is then given by (3.10) and (4.31),
$$\eqalign{
{\partial  \over {\partial t_{-1}}}\left( {\matrix{{u_{-1}}\cr
{u_0}\cr
}} \right)\,&=\,\left( {\matrix{0&1\cr
{1 / u_{-1}}&0\cr
}} \right)\,{\partial  \over {\partial t_0}}\,\left( {\matrix{{u_{-1}}\cr
{u_0}\cr
}} \right) \cr
 & {} \cr
&=\,{\partial  \over {\partial t_0}}\left( {\matrix{0&1\cr
1&0\cr
}} \right)\,\;\nabla H_0^{-1} , \cr} \eqno (8.4)
$$
where $\nabla H_0^{-1} := (\p H_0^{-1}/ \p u_{-1}, \p H_0^{-1}/ \p u_0)^T$, and
the hamiltonian function $H_0^{-1}$ is given by (4.16b),
$$H_0^{-1} = <\phi^{-1}>_0 :={\p F_0 \over \p u_{-1}}
= {1 \over 2} u_0^2 + u_{-1} ( \log u_{-1} -1).   \eqno (8.5)
$$
Here the universal free energy $F_0$ can be found as follows:
With the hamiltonian $H_0^0 = <\phi^0>_0 =\p F_0/\p u_0 = u_0u_{-1}$
for the identity flow, we obtain
$$
F_0 = {1 \over 2} u_0^2u_{-1} + {1 \over 2} u_{-1}^2
\left( \log u_{-1} -{3 \over 2} \right).   \eqno (8.6)
$$
The hamiltonian for the descendant field $H_1^{-1}$ of (5.11) becomes
$$
H_1^{-1} = <\sigma(\phi^{-1})>_0 = {1 \over 6} u_0^3 + u_0u_{-1}
( \log u_{-1} -1 ).  \eqno (8.7)
$$
The 2-point functions $<\phi^i \phi^j>$ then become
$$\eqalign{
& <\phi^0\phi^0>  = u_{-1}, \quad  <\phi^0\phi^{-1}>  = u_0, \quad
<\phi^{-1}\phi^{-1}> = {\rm log} \  u_{-1},  \cr
& <\sigma_1(\phi^0)\phi^0>   = u_0u_{-1}, \quad
<\sigma_1(\phi^0)\phi^{-1}>  = u_{-1} + {1 \over 2} u_0^2, \cr
& <\sigma_1(\phi^{-1})\phi^0>   = u_{-1}(\log u_{-1} - 1) + {1 \over 2} u_0^2,
\quad
<\sigma_1(\phi^{-1})\phi^{-1}>  = u_0 \log u_{-1} , \cr
& <\sigma_1(\phi^0)\sigma_1(\phi^0)>  = {1 \over 2}u_{-1}^2 + u_{-1}u_0^2, \cr
& <\sigma_1(\phi^0)\sigma_1(\phi^{-1})> = {1 \over 3}u_0^3 + u_{-1}u_0
 \log u_{-1} , \cr
& <\sigma_1(\phi^{-1})\sigma_1(\phi^{-1})>  = u_0^2 \log u_{-1} +
u_{-1} \{ (\log u_{-1})^2 - 2 \log u_{-1} + 2 \}, \cr }  \eqno (8.8)
$$
which are put in the string equation (6.8),
$$\eqalignno{
{\tilde t}_0 & +<\phi^0\phi_0>{\tilde t}_{1,0} + <\phi^{-1}\phi_0>
{\tilde t}_{1,-1}  \cr
& +
 <\sigma_1(\phi^0)\phi_0>{\tilde t}_{2,0} +
 <\sigma_1(\phi^{-1})\phi_0>{\tilde t}_{2,-1} \cdots  = 0,  & (8.9a) \cr
& {} \cr
{\tilde t}_{-1} & +<\phi^0\phi_{-1}>{\tilde t}_{1,0}
 + <\phi^{-1}\phi_{-1}>{\tilde t}_{1,-1}  \cr
& +
 <\sigma_1(\phi^0)\phi_{-1}>{\tilde t}_{2,0} +
 <\sigma_1(\phi^{-1})\phi_{-1}>{\tilde t}_{2,-1}\cdots  = 0.  & (8.9b) \cr}
$$

\medskip

As we mentioned in Section 2, we can set either $a) \ t_0=t_{\cal P}$ or
$b) \ t_{-1}=t_{\cal P}$.
The case of $a)$ corresponds to the dKP equation in the form of (8.4), while
in the case of b) the dKP equation is rewritten in the form (3.15)
with $\alpha_0 = -1$,
$$\eqalign{
{\partial  \over {\partial t_0}}\left( {\matrix{{\tilde u_{-1}}\cr
{\tilde u_0}\cr
}} \right)\,&=\,\left( {\matrix{0&e^{\tilde u_0}\cr
1&0\cr
}} \right)\,{\partial  \over {\partial t_{-1}}}\,\left( {\matrix{
{\tilde u_{-1}}\cr
{\tilde u_0}\cr
}} \right) \cr
&  {} \cr
&=\,{\partial  \over {\partial t_{-1}}} \left( {\matrix{0&1\cr
1&0\cr
}} \right)\,\;{\tilde \nabla} {\tilde H_0^0} ,
 \cr}
\eqno (8.10)
$$
where we define $\tilde u_{\alpha} =<\phi^{\alpha_0}\phi_\alpha>$ in (4.4) with
$\alpha_0 =-1$,
$${\tilde u_0} := <\phi^{-1}\phi_0> = \log u_{-1}, \quad
{\tilde u_{-1}} := <\phi^{-1}\phi_{-1}> = u_0.   \eqno (8.11)
$$
The hamiltonian function ${\tilde H}_0^0$ in (8.10) is given by
$$
{\tilde H_0^0} = <{\tilde \phi^0}>_{-1} := {\p F_{-1} \over \p {\tilde u_0}} =
{1 \over 2} {\tilde u_{-1}}^2 + e^{\tilde u_0},    \eqno (8.12)
$$
with the free energy $F_{-1}$ in (4.15b),
$$
F_{-1} = {1 \over 2} {\tilde u_{-1}}^2{\tilde u_0} + e^{\tilde u_0} .
 \eqno (8.13)
$$
In particular, note the universality of the 2-point function (4.14),
$$
<\phi^{\alpha}\phi^{\beta}>_0 := {\p^2 F_0 \over \p u_{\alpha} \p u_{\beta}} =
{\p^2  F_{-1} \over \p {\tilde u}_{\alpha} \p {\tilde u}_{\beta}} :=
<{\tilde \phi}^{\alpha}{\tilde \phi}^{\beta}>_{-1},  \eqno (8.14)
$$
where $\tilde \phi^{-1}=1$ and $\tilde \phi^0 = p-u_0 = e^{\tilde u_0} (p -
\tilde u_{-1})^{-1} \ ({\rm mod} \ W')$.
The string equations corresponding to these cases become as follows:

\medskip

For the case of a), choosing $C_{N,\alpha} = - \delta_{N,1}\delta_{\alpha,0}$
in (6.8),
$$\eqalignno{
 u_0 & =  t_{\cal P} + t_{1,0} u_0 + t_{1,-1}{\rm log} u_{-1} + \cdots ,
  & (8.15a) \cr
  u_{-1} & =  t_{\cal Q } + t_{1,0} u_{-1} + t_{1,-1} u_0 + \cdots ,
   & (8.15b) \cr }
$$
where $t_{\cal P} = t_{-1}$ is the primary coupling of ${\cal P}=
\phi^0 = \phi_{-1}$.  The free energy (6.14) for the
flat solution, say $F(t_{\cal P}, t_{\cal Q})$, is given by $F_0$ in (8.6)
with the substitution $u_0=t_{\cal P},\  u_{-1}=t_{\cal Q}$.

\medskip

For the case of $b)$, we have with the choice
$C_{N,\alpha} = - \delta_{N,1}\delta_{\alpha,-1}$ ,
$$\eqalignno{
 {\tilde u_{-1}} & =  t_{\cal P} + t_{1,0} e^{\tilde u_0} + t_{1,-1}
{\tilde u}_{-1}  \cdots ,   & (8.16a) \cr
  {\tilde u_0} & =  t_{\cal Q} + t_{1,0} {\tilde u_{-1}} + t_{1,-1}
{\tilde u}_0   \cdots .   & (8.16b) \cr }
$$
The free energy for the flat solution is then given by $F_{-1}$  in (8.13) with
${\tilde u_{-1}} = t_{\cal P}, \ {\tilde u_0} = t_{\cal Q}$.  This is the
 CP$^1$-model discussed in [DW, EY].
 As an example including a gravitational coupling, we
calculate the free energy (6.14) for the case with nonzero
$t_{1,-1} := t_{1, {\cal P}}$ [EHY]:  From (8.16), we first have
$$
\tilde u_{-1} = {t_{\cal P} \over 1 - t_{1,{\cal P}}},  \quad
\tilde u_0 = {t_{\cal Q} \over 1 - t_{1,{\cal P}}}.  \eqno (8.17)
$$
Then from (6.14) we obtain the same result as in [EHY],
$$
F(t_{\cal P}, t_{\cal Q}, t_{1, {\cal P}}) = {1 \over 2} {t_{\cal P}^2
t_{\cal Q} \over 1 - t_{1, {\cal P}}}  + ( 1 - t_{1, {\cal P}})^2
e^{t_{\cal Q} \over 1 - t_{1,{\cal P}}}.  \eqno (8.18)
$$
Here the point is that our derivation of the free energy is totally algebraic.

\vskip 0.3cm

\noindent
{\it $W_{2,0}$-model} :

The W potential is given by
$$W={1 \over 3} p^3 + v_1 p + v_0 .    \eqno (8.19) $$
where $v_0, v_1$ are related to the universal coordinates $u_0, u_1$ as
$$ v_0 = u^1 = u_0, \quad \quad v_1 = u^0 = u_1.    \eqno (8.20)$$
The flow equation for $t_1$ is given by
$$
{\partial  \over {\partial t_1}}\left( {\matrix{{u_1}\cr
{u_0}\cr
}} \right)\,=\,\left( {\matrix{0&1\cr
{-u_1}&0\cr
}} \right)\,{\partial  \over {\partial t_0}}\,\left( {\matrix{{u_1}\cr
{u_0}\cr
}} \right).   \eqno (8.21)
$$
In this model, we obtain the following two flat solutions and the corresponding
free energies:

a) With $u_\alpha=<\phi^0\phi_\alpha>$ for $\alpha=0$ and $1$,
$$F_0 = {1 \over 2}u_1u_0^2 - {1 \over 24} u_1^4,   \eqno (8.22)$$

b) With ${\tilde u}_0 := <\phi^1\phi_0> = - u_1^2 /2, \, \,
{\tilde u}_1 := <\phi^1\phi_1> = u_0$,
$$F_1 = {1 \over 2} {\tilde u}_0{\tilde u}_1^2 +
{1 \over 15} (-2{\tilde u}_0)^{5/2}.   \eqno (8.23)
$$
Note in this case that $F_1$ includes an algebraic singularity
 which indicates a critical phenomena.

\vskip 0.3cm

\noindent
{\it $W_{0,2}$-model}:

The W potential is given by
$$
W = p + {v_{-1} \over p-s} + {v_{-2} \over 2(p-s)^2}.  \eqno (8.24)
$$
with the relation,
$$
v_{-1} = u^0 = u_{-1}, \quad (v_{-2})^{1/2} = u^{-2} = u_{-2}, \quad
s = u^{-1} = u_0.   \eqno (8.25)
$$
The flow equations in the hierarchy are
$$\eqalignno{
{\partial  \over {\partial t_{-1}}}\left( {\matrix{{u_{-1}}\cr
{u_0}\cr
{u_{-2}}\cr
}} \right)\;&=\,\left( {\matrix{0&1&0\cr
0&0&{1/ u_{-2}}\cr
{1/ u_{-2}}&0&{-u_{-1}/ u_{-2}^2}\cr
}} \right)\;{\partial  \over {\partial t_0}}\left( {\matrix{{u_{-1}}\cr
{u_0}\cr
{u_{-2}}\cr
}} \right) \cr
 & {} \cr
&={\partial  \over {\partial t_0}}\left( {\matrix{0&1&0\cr
1&0&0\cr
0&0&1\cr
}} \right)\;\nabla H_0^{-1} , & (8.26a) \cr
{} \cr
{\partial  \over {\partial t_{-2}}}\left( {\matrix{{u_{-1}}\cr
{u_0}\cr
{u_{-2}}\cr
}} \right)\;&=\,\left( {\matrix{0&0&1\cr
{1/ u_{-2}}&0&{-u_{-1}/ u_{-2}^2}\cr
{-u_{-1}/ u_{-2}^2}&1&{u_{-1}^2/ u_{-2}^3}\cr
}} \right)\;{\partial  \over {\partial t_0}}\left( {\matrix{{u_{-1}}\cr
{u_0}\cr
{u_{-2}}\cr
}} \right) \cr
& {} \cr
&={\partial  \over {\partial t_0}}\left( {\matrix{0&1&0\cr
1&0&0\cr
0&0&1\cr
}} \right)\;\nabla H_0^{-2} , & (8.26b) \cr}
$$
where the hamiltonian functions are given by
$$\eqalignno{
H_0^{-1} & = <\phi^{-1}>_0 = {\p F_0 \over \p u_{-1}} = {1 \over 2} u_0^2 +
u_{-1} \log u_{-2}, & (8.27a) \cr
& {} \cr
H_0^{-2} & = <\phi^{-2}>_0 = {\p F_0 \over \p u_{-2}} = u_0u_{-2} + {1 \over 2}
{u_{-1}^2 \over u_{-2}}.  & (8.27b) \cr}
$$
Here the free energy $F_0$ is given by
$$
F_0 = {1 \over 2} u_0^2u_{-1} + {1 \over 2}u_{-2}^2u_0 +
{1 \over 2} u_{-1}^2 \log u_{-2}.   \eqno (8.28)
$$
In this model, there are three flat solutions:

\medskip

a) With the choice $C_{N,\alpha} = - \delta_{N,1}\delta_{\alpha,0}$, we
obtain the universal
free energy $F_0$ given by (8.28).  The free energy $F(t_{\alpha}: \alpha =
0, -1, -2)$ in (6.14) is then given by $F_0$ with $u_{\alpha} = t_{\alpha}$, in
particular, $t_0 = t_{\cal P}$.

\medskip

b) With $C_{N,\alpha} = -\delta_{N,1}\delta_{\alpha,-1}$,
$$F_{-1} = {1 \over 2} {\tilde u}_{-1} {\tilde u}_{-2}^2 + {1 \over 2}
{\tilde u}_0{\tilde u}_{-1}^2 - {1 \over 24} {\tilde u}_{-2}^4
+ {\tilde u}_{-2} e ^{{\tilde u}_0},   \eqno (8.29)
$$
where the new variables ${\tilde u}_{\alpha}, \ \alpha = 0, -1, -2$  are
defined by
$$\eqalign{
{\tilde u}_{-1} &:=<\phi^{-1}\phi_{-1}> = u_0, \cr
{\tilde u}_0 &:= <\phi^{-1}\phi_0> =\log u_{-2}, \cr
 {\tilde u}_{-2} &:= <\phi^{-1}\phi_{-2}> = u_{-1} / u_{-2}. \cr } \eqno (8.30)
$$
Note here that $<\phi^{\alpha}\phi^{\beta}> =
<{\tilde \phi}^{\alpha}{\tilde \phi}^{\beta}>_{-1} = \p^2 F_{-1} /
\p {\tilde u}_{\alpha} \p {\tilde u}_{\beta}$, and the free energy is given by
 $F(t_{\alpha}) = F_{-1}({\tilde u}_{\alpha} = t_{\alpha})$ with $t_{-1} =
 t_{\cal P}$.

\medskip

c) With $C_{N,\alpha} = -\delta_{N,1}\delta_{\alpha,-2}$,
$$F_{-2} = {1 \over 6} {\tilde u}_{-1} {\tilde u}_0^3 + {1 \over 6}
{\tilde u}_{-2}^3 +
{\tilde u}_0{\tilde u}_{-1}{\tilde u}_{-2} + {1 \over 2} {\tilde u}_{-1}^2
\left(\log {\tilde u}_{-1} - {3 \over 2} \right),   \eqno (8.31)
$$
where the new variables are
$$\eqalign{
{\tilde u}_{-1} &:=<\phi^{-2}\phi_{-1}> = u_{-2}, \cr
{\tilde u}_0 &:= <\phi^{-2}\phi_0> =u_{-1} / u_{-2}, \cr
 {\tilde u}_{-2} &:= <\phi^{-2}\phi_{-2}> =
u_0 - {1 \over 2} ( u_{-1} / u_{-2} )^2.  \cr } \eqno (8.32)
$$
The free energy is given in the same way as before, and $t_{-2} = t_{\cal P}$.

\vskip 0.3cm

\noindent
{\it $W_{1,1}$-model} :

The W potential is
$$W={p^2 \over 2} + v_0 + {v_{-1} \over p-s},   \eqno (8.33)$$
where
$$
v_0 = u^0 = u_0, \quad v_{-1} = u^1 = u_{-1}, \quad s = u^{-1} = u_1.
 \eqno (8.34) $$
This model has been recently studied in [Du2, KO].  Here we show that the flat
 solutions of this model coincide with those of the $W_{0,2}$-model just
  discussed (the equivalent pair discussed in Section 7).
With (8.33), the flow equations (3.10) are
$$\eqalignno{
{\partial  \over {\partial t_{1}}}\left( {\matrix{{u_{-1}}\cr
{u_0}\cr
{u_{1}}\cr
}} \right)\;&=\,\left( {\matrix{{u_1}&0&{u_{-1}}\cr
1&0&0\cr
0&1&{u_{1}}\cr
}} \right)\;{\partial  \over {\partial t_0}}\left( {\matrix{{u_{-1}}\cr
{u_0}\cr
{u_{1}}\cr
}} \right),   & (8.35a) \cr
{} \cr
{\partial  \over {\partial t_{-1}}}\left( {\matrix{{u_{-1}}\cr
{u_0}\cr
{u_{1}}\cr
}} \right)\;&=\,\left( {\matrix{0&1&{u_{1}}\cr
0&0&1\cr
{1 /u_{-1}}&0&0\cr
}} \right)\;{\partial  \over {\partial t_0}}\left( {\matrix{{u_{-1}}\cr
{u_0}\cr
{u_{1}}\cr
}} \right) . & (8.35b) \cr }
$$
The three flat solutions are as follows:

\medskip

a) With $C_{N,\alpha} = -\delta_{N,1}\delta_{\alpha,0}$,
$$
 F_0 = {1 \over 6}  u_{-1} u_1^3 + {1 \over 6}
u_0^3 +
 u_1 u_0 u_{-1} + {1 \over 2} u_{-1}^2
\left(\log  u_{-1} - {3 \over 2} \right).  \eqno (8.36)
$$
This gives the same free energy $F(t_{\cal P}, t_{{\cal Q}_1}, t_{{\cal Q}_2})$
as that in the case c) of the $W_{0,2}$-model, if we identify the variables
$u_{\alpha}$ in this model with ${\tilde u}_{\beta}, \ \beta = -(\alpha+2) \
({\rm mod} \ 3)$, in the $W_{0,2}$-model as
$$
t_{\cal P} := u_0 =  {\tilde u}_{-2}, \quad
t_{{\cal Q}_1} := u_1 =  {\tilde u}_0, \quad
t_{{\cal Q}_2} := u_{-1} =  {\tilde u}_{-1}. \quad  \eqno (8.37)
$$

\medskip

b) With $C_{N,\alpha} = -\delta_{N,1}\delta_{\alpha,-1}$,
$$F_{-1} = {1 \over 2} {\tilde u}_{-1} {\tilde u}_{0}^2 + {1 \over 2}
{\tilde u}_1{\tilde u}_{-1}^2 - {1 \over 24} {\tilde u}_{0}^4
+ {\tilde u}_{0} e ^{{\tilde u}_1},   \eqno (8.38)
$$
where
$$\eqalign{
{\tilde u}_{-1} &:=<\phi^{-1}\phi_{-1}> = u_0 + u_1^2 /2, \cr
{\tilde u}_0 &:= <\phi^{-1}\phi_0> =u_1, \cr
 {\tilde u}_{1} &:= <\phi^{-1}\phi_{1}> = \log u_{-1}. \cr } \eqno (8.39)
$$
The free energy is the same as in the case b) of the $W_{0,2}$-model.

\medskip

c) With $C_{N,\alpha}=- \delta_{N,1}\delta_{\alpha,1}$,
$$F_1 = {1 \over 2} {\tilde u}_{-1} {\tilde u}_1^2 + {1 \over 2}
{\tilde u}_1{\tilde u}_0^2 + {1 \over 2} {\tilde u}_{-1}^2
\log {\tilde u}_0 ,   \eqno (8.40)
$$
where
$$\eqalign{
{\tilde u}_{-1} &:=<\phi^1\phi_{-1}> = u_1u_{-1}, \cr
{\tilde u}_0 &:= <\phi^1\phi_0> =u_{-1} , \cr
 {\tilde u}_1 &:= <\phi^1\phi_1> = u_0 + u_1^2 / 2.  \cr } \eqno (8.41)
$$
The free energy is again the same as in the case a) of the $W_{0,2}$-model.

\vfill\eject

\noindent
{\bf Appendix A: Calculus including log-terms}

\vskip 0.6cm

Here we give the explicit calculations of quantities including log-terms,
for example, the generators $Q_N^{-1}$ in (5.7), and the 1-point function
(4.16b) of the corresponding fields $\sigma_N(\phi^{-1})$.

\medskip

We first define the projection symbols $[\cdot]_+$ and $[\cdot]_-$ in (2.5)
in terms of the contour integrals.  Let $f(p) \in {\C}[p, (p-s)^{-1}]$, that
 is, $f(p)$ is a rational function given by
$$
f(p) = \sum_{k \ge 0} a_k  p^k  +  \sum_{l \ge 1} {b_l \over (p-s)^l}.
 \eqno (A.1)
$$
Then the $+$ and $-$ projections are defined by
$$\eqalign{
[f(p)]_+  &:= \sum_{k \ge 0} a_k  p^k , \cr
[f(p)]_-  &:= \sum_{l \ge 1} {b_l \over (p-s)^l}. \cr }  \eqno (A.2)
$$
The coefficients $a_l$ and $b_l$ are obtained by
$$\eqalign{
a_k &= \mi \left[{f(p) \over p^{k+1}} \right] := {1 \over 2\pi i}
\oint_{C_\infty } {f(p) \over p^{k+1}} \ dp \ ,  \cr
& {} \cr
b_l &= \ms [f(p) (p-s)^{l-1}] := {1 \over 2\pi i}
\oint_{C_s } {f(p) (p-s)^{l-1} \ dp} , \cr }  \eqno (A.3)
$$
where $C_{\infty}$ and $C_s$ are the contoures oriented in the anti-clockwise
about $p=\infty$ and $p=s$, respectively.  Then from (A.2) and (A.3), we have
$$\eqalign{
[f(p)]_+ &= {1 \over 2\pi i}
\oint_{C_\infty } {f(z) \over z-p}  dp , \cr
& {} \cr
[f(p)]_- &= - {1 \over 2\pi i}
\oint_{C_s} {f(p) \over z-p}  dp . \cr }  \eqno (A.4)
$$
Note that the point $p$ in (A.4) locates between $C_\infty$ and $C_s$.
Eqs.(A.4) give explicit formulae for the projections of the rational functions
in ${\C}[ p, (p-s)^{-1}]$.

\medskip

Now let us consider the case including log-terms,
$\log \lambda$ and $\log \mu$.
In this case, one has to modify $C_s$ in (A.4) to $\tilde C_s$ which
is taken to surround a branch cut between $p=s$ and $p=\infty$.
The main idea of defining these terms is to regularize the log-terms by adding
and subtracting the singular parts of these terms.
For the formula $Q_N^{-1}$
in (5.7) is then calculated as follows:  We first write
 $\log \lambda = \log (\lambda /p) + \log p$, and  $\log \mu
= \log [\mu (p-s)] - \log (p-s)$.  Then  we have
$$\eqalign{
Q_N^{-1} = &\left[ {W^N \over N!}( \log \lambda - c_{N}) \right]_+  -
\left[ {W^N \over N!}( \log \mu - d_{N}) \right]_-  \cr
& {} \cr
= &\left[ {W^N \over N!} \left( \log {\lambda \over p} - c_{N} \right)
\right]_+ -
\left[ {W^N \over N!} \left( \log [\mu (p-s)] - d_{N} \right) \right]_- \cr
& {} \cr
& + \left[ {W^N \over N!} \log p \right]_+ +
\left[ {W^N \over N!} \log (p-s) \right]_- . \cr }  \eqno (A.5)
$$
The first two terms are well defined and give polynomials in
 ${\C}[ p, (p-s)^{-1}]$, while
the last two terms may be computed by deforming the contour
in the integrals (A.4),
i.e. with $C_s \to {\tilde C_s}$
$$\eqalign{
 &\left[ {W^N \over N!} \log p \right]_+ +
\left[ {W^N \over N!} \log (p-s) \right]_-  \cr
& {} \cr
&:=   {1 \over 2\pi i N!}
\oint_{C_\infty } {W(z)^N \log z \over z-p}  dz -
{1 \over 2\pi i N!}
\oint_{\tilde C_s} {W(z)^N \log (z-s) \over z-p}  dz  \cr
& {} \cr
&=  {1 \over 2\pi i N!}
\oint_{C_\infty } {W(z)^N \over z-p} \log \left( {z \over z-s} \right)
 dz  + {W(p)^N \over N!} \log (p-s).  \cr }  \eqno (A.6)
$$
Thus $Q_N^{-1}$ includes the log-term, and is calculated as
$$\eqalign{
Q_N^{-1} = &\left[ {W^N \over N!}\left( \log  {\lambda \over p-s}
- c_{N} \right) \right]_+  -
\left[ {W^N \over N!} \left( \log [\mu (p-s)] - d_{N} \right) \right]_-  \cr
& {} \cr
&+ {W(p)^N \over N!} \log (p-s) . \cr }  \eqno (A.7)
$$
In particular, we see from (A.7) that $Q^{-1} = Q_0^{-1} = \log (p-s)$
in (2.5).
Note here that the fields $\sigma_N(\phi^{-1}) = {\p Q_N^{-1}/ \p p}$ are all
well-defined as elements in the ring (2.10),
${\cal R} = {\C}[ p, (p-s)^{-1}] / W'(p)$.

\medskip

In a similar way as above, the residue formula including the log-terms can be
also explicitly expressed by the contour integrals. Let
$f(p) \in {\C}[p, (p-s)^{-1}]$.
 We then want to give a precise meaning
of the quantity, e.g. (5.11),
$$
\mi [ \ f(p)   \log \lambda \ ] + \ms [ \ f(p)  \log \mu \ ].   \eqno (A.8)
$$
By the regularizations for $\log \lambda$ and $\log \mu$, we obtain
$$\eqalign{
&\mi \left[ \ f(p) \log {\lambda \over p} \right]
+ \ms \left[ \ f(p) \log [\mu (p-s)] \right]
+ {1 \over 2\pi i}
\oint_{C_\infty} f(p)[\log p - \log (p-s) ]  dp \cr
& {} \cr
&= \mi \left[ f(p) \log {\lambda \over p-s} \right] +
\ms [f(p) \log \{ \mu (p-s) \}]. \cr }  \eqno (A.9)
$$
Note that the contour integral in (A.9) gives $\int_0^s [f(p)]_+ dp$
which was previously obtained in [AK], i.e.
$$
\oint_{C_\infty} f(p)[\log p - \log (p-s) ]  dp =
\int_0^s dz \ \oint_{C_\infty} {f(p) \over p-z} \ dp.
$$

\vskip 1.5cm

\noindent
{\bf Appendix B:  The dKP hierarchy}

\vskip 0.5cm

In this Appendix, we give a brief summary of the dKP theory
as a quasi-classical limit of the KP
theory.  For a simplicity, we consider here the original form of
the KP hierarchy, and not a constrained
hierarchy discussed recently in [BX, AFNV, OS], which is directly related
to our dKP hierarchy in the dispersionless limit.
The KP theory without constraint may be formulated
as follows:

\vskip 0.3cm

Let $L$ be a formal pseudo-differential operator given by
$$
L = \p  +  \sum_{i=0}^\infty A^i \p^{-(i+1)},   \eqno (B.1)
$$
where $A^i=A^i (X,T_1,T_2,\cdots)$, the symmbol $\p$ implies the derivative
with respect to $X$, and $\p^{-1}\p = \p \p^{-1} = 1$.  The operation with
$\p^i$ is given by a generalized Leibnitz rule,
$$\partial ^iF=\sum\limits_{j=0}^\infty  {\left( {\matrix{i\cr
j\cr
}} \right)}{{\partial ^jF} \over {\partial X^j}}\partial ^{i-j}\quad,\quad
   i \in \Z.  \eqno (B.2) $$
The KP hierarchy is then defined by the so-called Lax formula,
$${{\partial L\over \p T_n}}
=[B_n\,,\,L]:=\;B_n\,L\,-\,\,
L\,B_n\;,\;\quad\,\;\ \ n=0,1,\cdots  \eqno (B.3)  $$
where the differential operator $B_n$ is given by the
differential part of $L^{n+1}/(n+1)$, denoted by
$$ B_n = {1 \over {n+1}} [ L^{n+1}]_+ \ .  \eqno (B.4)
$$
The hierarchy (B.3) is also given by the compatibility conditions of
the following linear equations for the wave function $\Psi (T_0,T_1,\cdots)$
 with $T_0=X$, and  ${\p \lambda \over \p T_n} = 0$;
$$\eqalignno{ L\Psi  & =\lambda \Psi , & (B.5a) \cr
& {} \cr
{{\partial \Psi } \over {\partial T_n}} & = B_n \Psi .  & (B.5b) \cr
}$$
Note that (B.5b) gives an iso-spectral deformation of the operator
$L$ in (B.5a). Then, the dKP hierarchy can be obtained from
the``quasi-classical" limit of the KP theory as follows [Ko2]:
Let $\h$ be a small parameter (the Plank constant),
and introduce the variables,
$$\eqalignno{
 t_n  & := {\h} T_n , \quad \quad \quad \quad {\rm for \ } \quad
 n = 0, 1, \cdots, & (B.6a) \cr
 a^i & = a^i (t_0, t_1, \cdots)   := A^i(T_0, T_1, \cdots) ,
 \quad {\rm for } \quad  i = 0, 1, \cdots,  & (B.6b) \cr }
$$
which lead to the replacement ${\p \over \p T_n}= {\h} {\p \over \p t_n}$.
Then write the wave function $\Psi$ in (B.5) to be the WKB form,
$$
\Psi(T_0,T_1, \cdots) = {\rm exp} \left\{{{1\over \h} S(t_0,t_1,\cdots)}
\right \}. \eqno (B.7)
$$
where the function S is called the action, and plays a fundamental
role of the dKP theory.  With (B.7) the quasi-classical limit leads to,
 for $i \in \Z$,
$${{{\partial_X} ^i\Psi } \over \Psi }= {\h}^i {{{\p _x}^i \Psi } \over \Psi}
\;\to \;p^i,\quad {\rm as} \quad \h\;\to \;0.
  \eqno (B.8)
$$
where $p$ is the momentum function defined by
$$
p = {\p S \over \p x} .  \eqno (B.9)
$$
{}From (B.8), eqs.(B.5) in the limit become
$$
\eqalignno{
& \lambda  = p + {a^0 \over p} + {a^1 \over p^2} + \cdots,
 & (B.10a)  \cr
 & {} \cr
&{\p p \over \p t_n} =  {\p Q^n \over \p x}, & (B.10b)     \cr}
$$
where $Q^n$ given by
$\mathop {lim}\limits_{\h \to \infty }\;[B_n\Psi / \Psi ]$
is the polynomial part  of $\lambda^{n+1} /(n+1)$ in $p$, and
we denote $Q^n = [\lambda^{n+1}/(n+1)]_+$ as in (2.5).
Note that (B.10) is the
Hamilton-Jacobi equation for the wave equation (B.5), and it defines the dKP
hierarchy.  In this formulation, $\lambda $ is considered to be a constant
given by the spectral parameter of $L$.  Note that the formulation we used
in the text is different from the one given here, that is, the momentum
function $p$ in the text is considered as a parameter instead of $\lambda$.
However, these formulations are
of course equivalent, and indeed they are connected as a
cannonical change of variables:
Namely consider the differential three-forms ( $\infty$-forms in general),
$$ d\lambda \wedge dp \wedge dx = dQ^n \wedge d\lambda \wedge dt_n ,
\eqno (B.11)
$$
which leads to both (B.10) and the dKP hierarchy in the form (2.4) by
considering the independent variables to be either $(\lambda, t_n, x)$ or
$(p, t_n, x)$, and comparing the coefficients of
$d\lambda \wedge dt_n \wedge dx$ or
$dp \wedge dt_n \wedge dx$.  Now it is clear from (B.10) that the function $p$
gives the conserved densities of the hierarchy (Theorem 3.1).  Also the
compatibility conditions among the flows in (B.10), which are now given by
$$
{\p Q^i \over \p t_j} = {\p Q^j \over \p t_i}\ .  \eqno (B.12)
$$
Note from (B.9) and (B.10b)
that $Q^i$ is written in the form with the action $S$,
$$ Q^i = {\p S \over \p t_i}, \quad \quad  {\rm for } \quad  n = 0, 1, \cdots .
 \eqno (B.13)
$$
Writing $Q^i$ in a Laurent series of $\lambda$, we have
$$
Q^i = {1 \over i+1} [\lambda^{i+1}]_+ := {1 \over i+1} \lambda^{i+1} -
\sum_{j=0}^{\infty} {1\over \lambda^{j+1}} G^{ij} .  \eqno (B.14)
$$
Here the coefficients $G^{ij}$  can be calculated by the residue form,
$$
G^{ij} = - {\mathop {res}\limits_{\lambda = \infty}} [ Q^i \lambda^j ] =
{1 \over j+1} \mi \left[ \lambda^{i+1} {\p Q^i \over \p p} \right], \eqno
(B.15)
$$
which is just (3.11a), and also shows $G^{ij} = G^{ji}$.  We then see that
the action $S$ can be written in terms of the free energy $F$,
$$
S = {\sum_{i=0}^\infty t_i {\lambda^{i+1} \over {i+1}}} -
   \sum_{j=0}^\infty {1 \over \lambda^{j+1}}{\p F \over  \p t_j} .
\eqno (B.16)
$$
Here the free energy is defined  by (4.2), $G^{ij} = \p^2 F / {\p t_i}{\p
t_j}$.
The existence of the free energy is a consequence of the integrability (B.12),
i.e. (3.14),
$$
{\p G^{ij} \over \p t_k} = {\p G^{kj} \over \p t_i}.  \eqno (B.17)
$$
Noticing the scale invariance of (B.17) under $t_i \to \epsilon t_i$, we see
that the functions $G^{ij}$ are  homogeneous functions of
degree zero, and so that the free energy $F$ is of degree two , i.e. (6.16)
[Kr1, TT2].   This implies
$$
\sum_{i=0}^\infty t_i {\p \over \p t_i} F = 2 F .   \eqno (B.18)
$$
Then taking the derivative of (B.18) with respect to $t_j$ leads to
$$
\sum_{i=0}^\infty t_i {\p^2 \over \p t_i \p t_j} F =
{\p F \over \p t_j}, \eqno (B.19)
$$
and, using (B.18) once again, we obtain the formula (6.14) of
the free energy, i.e.
$$
F = {1 \over 2} \sum_{{i,j}=0}^\infty t_it_j G^{ij}.
  \eqno (B.20)
$$
Note also that using the formula (B.15)
the free energy $F$ can be written in the form [Kr1],
$$
F = {1 \over 2}  \mi \left[ S^+ {\p S_+ \over \p p} \right],   \eqno (B.21)
$$
where $S^+ := \sum_{i=0}^{\infty} t_i \lambda^{i+1}/(i+1)$, and
$S_+ = \sum_{i=0}^{\infty} t_i Q^i $.

\vfill\eject

\noindent
{\bf References}

\vskip 0.4cm

\item{[AK]} S. Aoyama, and Y. Kodama: Mod. Phys. Lett. A9 (1994) 2481.
\item{[ANPV]} H. Aratyn, E. Nissimov, S. Pacheva, and I. Varysburg: Phys. Lett.
B294 (1992)167.
\item{[BDKS]} E. Brezin, M. Douglas, V. Kazakov, and S. Shenker: Phys. Lett.
B237 (1990) 43.
\item{[BDSS]} T. Banks, M. Douglas, N. Seiberg and S. Shenker, Phys. Lett.
B238 (1990) 279.
\item{[BKa]} E. Brezin, and V. Kazakov: Phys. Lett. B236 (1990) 144.
\item{[BKo]} A. M. Block, and Y. Kodama: SIAM J. Appl. Math. 52 (1992) 909.
\item{[BX]} L. Bonora, and C. S. Xiong: Int. J. Mod. Phys. A8 (1993) 2973;
Nucl. Phys. B405 (1993) 191; Phys. Lett. B317 (1993) 329.
\item{[CGM]} C. Crnkovic, P. Ginsparg and G. Moore, Phys. Lett. B237 (1990)
196.
\item{[Da]} F. David, Mod. Phys. Lett. A3 (1988) 1651.
\item{[Do]} M. Douglas: Phys. Lett. B238 (1990) 176.
\item{[Du1]} B. Dubrovin: Comm. Math. Phys. 145(1992)195; Nucl. Phys. B379
(1992) 627.
\item{[Du2]} B. Dubrovin: Geometry of 2D topological field theory,
SISSA-84/94/FM, hep-th/9407018.
\item{[DK]} J. Distler and H. Kawai, Nucl. Phys. B321 (1989) 509.
\item{[DS]} M. Douglas, and S. Shenker: Nucl. Phys. B335 (1990) 635.
\item{[DVV1]} R. Dijkgraaf, H. Verlinde and E. Verlinde: Nucl. Phys. B352
 (1991) 59.
\item{[DVV2]} R. Dijkgraaf, H. Verlinde and E. Verlinde: Nucl. Phys. B348
 (1991) 435.
\item{[DW]} R. Dijkgraaf and E. Witten: Nucl. Phys. B342 (1990) 486.
\item{[EHY]} T. Eguchi, K. Hori, and S. -K. Yang: Topological $\sigma$-model
and large-N matrix model, UT-700, hep-th/9503017.
\item{[EKYY]} T. Eguchi, H. Kanno, Y. Yamada and S.-K. Yang: Phys. Lett.
B305 (1993) 235.
\item{[EY]} T. Eguchi, and S.-K. Yang: Mod. Phys. Lett. A9 (1994) 2893.
\item{[EYY1]} T. Eguchi, Y. Yamada and S.-K. Yang: Mod. Phys. Lett. A8 (1993)
1627.
\item{[EYY2]} T. Eguchi, Y. Yamada and S.-K. Yang: On the Genus Expansion
in the Topological String Theory, UTHEP-275, hep-th/9405106.
\item{[FGZ]} P. Di Francesco, P. Ginsparg, and J. Zinn-Justin: 2D Gravity
and Random Matrices, LA-UR-93-1722, SPhT/93-061, hep-th/9306153.
\item{[FK]} P. D. Francesco, and D. Kutasov,  Nucl. Phys. B342 (1990) 589.
\item{[FKN]} M. Fukuma, H. Kawai, and R. Nakayama:  Int. J. Mod. Phys. A6
(1991) 1385.
\item{[GGPZ]} P. Ginsparg, M. Goulian, M. R. Plesser, and J. Zinn-Justin:
Nucl. Phys. B342 (1990) 539.
\item{[GM]} D. Gross, and A. Migdal:  Nucl. Phys.
B340 (1990) 333.
\item{[Ka]} V. Kazakov: Mod. Phys. Lett. A4 (1989) 2125.
\item{[Ko1]} Y. Kodama: Phys. Lett. A147 (1990) 477.
\item{[Ko2]} Y. Kodama: Prog. Theor. Phys. Supp. 94  (1988) 184; Phys Lett.
A129 (1988) 223.
\item{[Kr1]} I. M. Krichever: Comm. Math. Phys. (1991) 415.
\item{[Kr2]} I. M. Krichever: Comm. Pure. Appl. Math. 47 (1994) 437.
\item{[KG]} Y. Kodama and J. Gibbons: Phys. Lett. A135 (1989) 167;
Integrability of the dispersionless KP hierarchy, Proceedings of the workshop
``Non-linear Processes in Physics" (World Scientific, 1990) 166.
\item{[KO]} H. Kanno, and Y. Ohta: Topological strings with scaling violation
and Toda lattice hierarchy, hep-th/9502029.
\item{[KPZ]} V.G. Knizhnik, A.M. Polyakov and A.B. Zamolodchikov, Mod. Phys.
Lett. A3(1988)819.
\item{[Lo]} A. Losev: Theor. Math. Phys. 95 (1993) 595.
\item{[LP]} A. Losev and L. Polyubin: On connection between topological
 Landau-Ginzburg gravity and integrable systems, hep-th/9305079.
\item{[OS]} W. Oevel, and W. Strampp: Comm. Math. Phys. 157 (1993) 51.
\item{[S]} I. A. B. Strachan: The Moyal bracket and the dispersionless
limit of the KP hierarchy, hep-th/9410048.
\item{[SV]} M. V. Saveliev, and A. M. Vershik:
Comm. Math. Phys. 126 (1989) 367.
\item{[T]} K. Takasaki: Lett. Math. Phys. 29 (1993) 111.
\item{[TT1]} K. Takasaki and T. Takebe: Lett. Math. Phys. 23 (1991) 205.
\item{[TT2]} K. Takasaki and T. Takebe: Int. J. Mod. Phys. A. Supp. (1992) 889.
\item{[VW]} C. Vafa and N.P. Warner: Phys. Lett. B218 (1989) 51;
\item{[W]} E. Witten: Nucl. Phys. B340 (1990) 281.
\item{[Z]} B. Zuber: On Dubrovin topological field theories,
SPht 93/147, hep-th/ 9312209.

\vfill\eject




\bye